\documentclass[twocolumn]{aastex631}

\usepackage{amsmath}
\usepackage{natbib}
\newcommand{\package}[1]{\textsl{#1}}

\newcommand{\msun}{\textrm{M}_\odot}

\renewcommand{\tablename}{Table}

\defcitealias{pearson19}{P19}

\shorttitle{GC streams in M31}
\shortauthors{Pearson, Clark et al.}

\begin{document}

\title{The \texttt{Hough Stream Spotter}: \\
A New Method for Detecting Linear Structure in Resolved Stars and Application to the Stellar Halo of M31} 

\author[0000-0003-0256-5446]{Sarah Pearson}\thanks{Hubble Fellow}
\affiliation{Center for Cosmology and Particle Physics, Department of Physics, New York University, 726 Broadway, New York, NY 10003, USA}
\affiliation{Center for Computational Astrophysics, Flatiron Institute, 162 5th Av., New York City, NY 10010, USA}

\author[0000-0002-7633-3376]{Susan E. Clark}
\affiliation{KIPAC, Stanford University, 2575 Sand Hill Road, Menlo Park, CA  94025, USA }

\author{Alexis J. Demirjian}
\affiliation{Georgetown University, 37th and O Streets, N.W., Washington DC, USA}

\author[0000-0001-6244-6727]{Kathryn V. Johnston}
\affiliation{Columbia University, 550 West 120th Street, New York City, NY 10027, USA}
\affiliation{Center for Computational Astrophysics, Flatiron Institute, 162 5th Av., New York City, NY 10010, USA}

\author[0000-0001-5082-6693]{Melissa K. Ness}
\affiliation{Columbia University, 550 West 120th Street, New York City, NY 10027, USA}

\author[0000-0003-2539-8206]{Tjitske K. Starkenburg}
\affiliation{Center for Interdisciplinary Exploration and Research in Astrophysics (CIERA) and \\Department of Physics and Astronomy, Northwestern University, 1800 Sherman Ave, Evanston, IL 60201, USA}

\author[0000-0002-7502-0597]{Benjamin F. Williams}
\affiliation{Department of Astronomy, Box 351580, University of Washington, Seattle, WA 98195, USA}

\author[0000-0002-3292-9709]{Rodrigo A. Ibata}
\affiliation{Observatoire astronomique de Strasbourg, Universit\'{e} de Strasbourg, CNRS, UMR 7550, 11 rue de l'Universit\'{e}, F-67000 Strasbourg, France}

\begin{abstract}\noindent 
Stellar streams from globular clusters (GCs) offer constraints on the nature of dark matter and have been used to explore the dark matter halo structure and substructure of our Galaxy.
Detection of GC streams in other galaxies would broaden this endeavor to a cosmological context, yet no such streams have been detected to date.
To enable such exploration, we develop the \texttt{Hough Stream Spotter} (\texttt{HSS}), and apply it to the Pan-Andromeda Archaeological Survey (PAndAS) photometric data of resolved stars in M31's stellar halo. We first demonstrate that our code can rediscover known dwarf streams in M31. We then use the \texttt{HSS} to blindly identify 27 linear GC stream-like structures in the PAndAS data. For each \texttt{HSS} GC stream candidate, we investigate the morphologies of the streams and the colors and magnitudes of all stars in the candidate streams. We find that the five most significant detections show a stronger signal along the red giant branch in color-magnitude diagrams (CMDs) than spurious non-stream detections. Lastly, we demonstrate that the \texttt{HSS} will easily detect globular cluster streams in future Nancy Grace Roman Space Telescope data of nearby galaxies. This has the potential to open up a new discovery space for GC stream studies, GC stream gap searches, and for GC stream-based constraints on the nature of dark matter.

\end{abstract}

\section{Introduction} \label{sec:intro}
More than 60 stellar streams have been detected in the Milky Way  \citep[MW;][]{mateu18}. These streams have been identified from a variety of search methods \citep[e.g.,][]{johnston96,grillmair95,Rockosi02,grillmair06,Shipp18, shih21}, and they have taught us crucial information about the mass distribution of dark matter \citep[e.g.,][]{koposov10,kuepper15,bovy16,bonaca18,malhan19,reino20} and the accretion history of our Galaxy \citep[e.g.,][]{newberg02,belokurov06,helmi18}. 
Thin stellar streams that emerge from globular clusters (GCs) are particularly useful, as their small physical scales and low velocity dispersion (only a few kilometers per second), make them sensitive to subtleties in potential properties that can be noticeable in the morphology of the streams alone. As GC streams are dynamically cold, i.e. their velocity dispersion is much smaller than their orbital velocity around their host galaxy, morphological and kinematic disturbances in the GC streams remain distinct and coherent for billions of years. 

GC stellar streams are more sensitive to gravitational disturbances from nearby low-mass dark matter subhalos than stellar streams emerging from dwarfs. These disturbances can create gaps and density fluctuations in the GC streams \citep[e.g.,][]{yoon11, carlberg12}. 
If we find indirect evidence of low-mass subhalos through gaps in streams, this can help rule out certain dark matter particle candidates \citep[e.g.,][]{bullock17}. 
In the MW, there are several examples of streams deviating from coherent structures \citep[e.g.,][]{bonaca19c,li20,bonaca21b}, but less than a handful of examples of streams with gaps \citep[e.g.,][]{oden01,price18,li20}.

GC stellar streams are also useful because their morphologies can constrain the global structure of our Galaxy, through the precession of the streams' orbits \citep[e.g.,][]{johnston2002,dehnen04,johnston2005,belokurov14,erkal16}. In  \cite{pearson15}, we showed that a specific triaxial dark matter halo that had been proposed to explain the 6D structure of the Sagittarius stream \citep{law10} could be ruled out using a constraint from the 2D morphology of the Palomar 5 (Pal 5) stream because its stream became ``fanned''. Such ``fanned'' streams exist near abrupt transitions between orbit families known as {\it separatrices} \citep{price16,yavetz21}. 
These results suggest that observations of thin streams map out smooth transitions in orbital properties supported by a potential, while disturbed stream morphologies (or the absence of streams altogether) may be used to identify separatrices between orbit families. While these effects from time-dependent perturbations and global structure will also apply to more massive, dwarf streams, the low velocity dispersion in the GC streams makes them particularly sensitive.

A common limitation to MW studies is cosmic variance -- that we are only studying one galaxy.  Extending the sample of thin GC streams to hundreds of galaxies is crucial if we want to 1) expand the sample of clear gaps in streams that could originate from interactions with dark matter subhalos, and 2) probe the dark matter mass distributions in more galaxies. 
Several dwarf galaxy streams have already been discovered in external galaxies \citep[e.g.,][]{mcconnachie2009,delgado10,denja16}. 
However, we still do not have clear evidence of any GC streams in galaxies other than the MW. GC streams are much fainter and thinner than dwarf stellar streams and are therefore harder to detect against diffuse backgrounds of stellar halos in external galaxies. Over the next decade, data from upcoming telescopes such as the Nancy Grace Roman Space Telescope  \citep[Roman, formerly WFIRST;][]{spergel15}, the Vera Rubin Observatory  \citep[VRO, formerly LSST;][]{laureijs11}, and Euclid \citep[][]{racca16} will reveal thousands of dwarf stellar streams in external galaxies, as well as a number of GC streams \citep{pearson19}. 

In this paper, we develop a new stream-finding code the \texttt{Hough Stream Spotter} (\texttt{HSS}: \citealt{Pearson2021})\footnote{Also available at \url{https://github.com/sapearson/HSS}.} which is designed to detect and quantify stream signatures in large data sets through a Hough transform \citep{hough62}. Our approach requires only the 2D plane of stellar positions as input, thus it can be applied across missions. 
For external galaxies, the number of data sets with 2D projections (i.e. sky positions) will greatly exceed those with any kinematic information (i.e. line-of-sight velocities), and there is no prospect of gathering the 6D phase-space maps that are attainable for the MW. Thus, there is a need for algorithms that can search blindly (or semi-blindly) through these large data sets and identify stream candidates, especially in noisy or background-confused data. 
The \texttt{HSS} fills this need as it is computationally efficient and simultaneously quantifies the linearity of stream candidates relative to the background.

\citet{pearson19} (hereafter \citetalias{pearson19}) showed that Roman, planned to launch in mid-2020s, will easily detect GC streams in resolved stars in galaxies out to at least 3.5 Mpc. More than 80\% of the galaxies within this volume are dwarfs \citep[see][]{kara19}, and many of these galaxies do not harbor molecular clouds, spiral arms, or bars, which can also produce gap signatures in streams and contaminate the subhalo signal \citep{amorisco16, pearson17, erkal17, banik19}. 
If we discover GC streams in external galaxies, which the \texttt{HSS} is set up to do, this provides exciting prospects for studying morphologies and stream gaps (i.e. dark matter subhalo populations) as a function of galactic radii \citep[see e.g.,][]{kimmel17} and environment in a large sample of host galaxies, which could help uncover the nature of dark matter. 
With the \texttt{HSS}, we can fully exploit our growing observational data sets and advance our understanding of how  thin  streams  might  constrain  dark matter distribution  and  properties  through {\it morphology alone}.

We  introduce  and  validate  our automated approach to stream-finding by applying the \texttt{HSS} to the Pan-Andromeda Archaeological Survey (PAndAS) stellar halo data \citep{mcconnachie2009,McConnachie18}\footnote{Obtained from R. Ibata, private communication.} where we first identify the known dwarf galaxy streams and subsequently do a blind search for new GC stream candidates. 
\citetalias{pearson19} showed that an old GC stream, scaled to have five to ten times more mass than the stream emerging from the MW globular cluster, Pal 5, would be detectable in the PAndAS data after applying a metallicity cut of [Fe/H] $< -1$.
More than 450 GCs have been detected in M31 to date \citep[][]{huxor14,caldwell16,mackey19}. This is more than three times the GC population in the MW, and this large difference likely arises from dissimilarities between the two spiral galaxies' accretion histories \citep[e.g.,][]{deason13,forbes18}. 
\citet{huxor14} searched for stellar streams surrounding the known globular clusters in M31 using Hubble Space Telescope (HST) data and did not detect any associated stellar streams. For the majority of the known GC streams in the MW, the progenitor has been fully disrupted \citep[e.g.,][]{balbinot18}. Examples of extended GC stellar streams with associated progenitors exist (e.g., Pal 5: \citealt{oden01}, NGC 5466: \citealt{grillmair06b}, $\omega$Cen: \citealt{ibata19}, Pal 13: \citealt{ship20}), and \citet{ibata21} recently reported 15 streams associated with known MW GCs in Gaia DR3 \citep{gaiadr3}. However, despite the fact that \citet{huxor14} did not find any GC streams with deep follow-up observations near the GCs, we might be able to detect GC streams in a blind search of the M31 stellar halo by running the \texttt{HSS}.

The paper is organized as follows: in Section \ref{sec:data} we describe the M31 PAndAS data. In Section \ref{sec:RHT}, we present our code.  
In Section \ref{sec:applypandas}, we describe how we apply our code to PAndAS data. In particular, we show how we treat the regions and mask out known objects (Section \ref{sec:input}), we optimize our code to search for GC streams in M31 (Section \ref{sec:optimize}), we 
demonstrate that the code easily detects known M31 dwarf streams (\ref{sec:dwarfs}), and we carry out completeness tests of our code using synthetic streams (Section \ref{sec:completeness}).
In Section \ref{sec:results}, we show the results of blindly running the \texttt{Hough Stream Spotter} on PAndAS data, present our GC stream candidates (Section \ref{sec:resrht}), and analyze the morphologies and color-magnitude diagrams (CMDs) of our stream candidates (Section \ref{sec:CMD}). We
discuss the implications of our results and comparisons to other stream-finding techniques in Section \ref{sec:newdiscussion}, and we review the future prospects of GC stream searches in external galaxies in Section \ref{sec:discussion}. We conclude in Section \ref{sec:conclusion}.

\begin{figure}[htp]
\centerline{\includegraphics[width=\columnwidth]{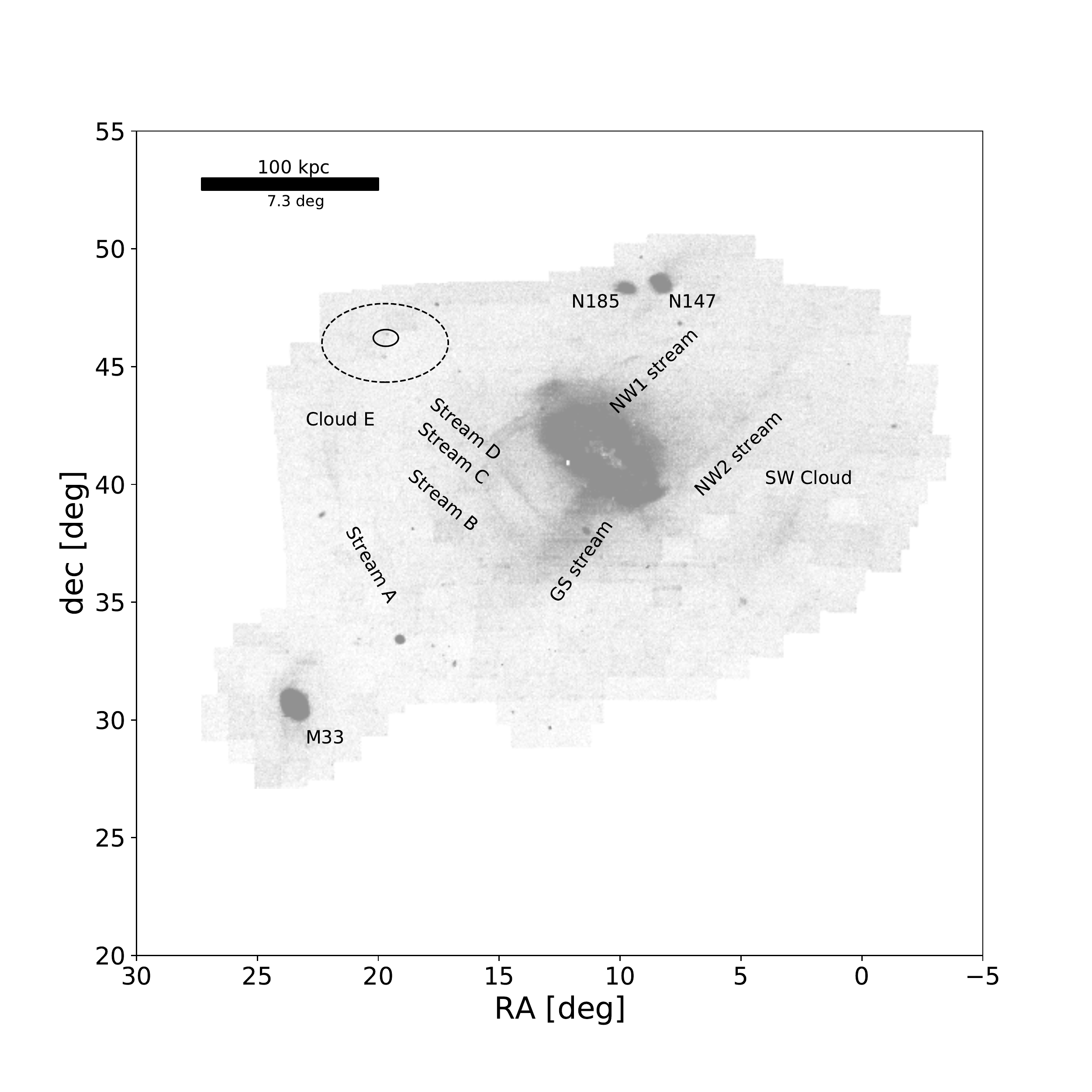}}
\caption{PAndAS observations of the Andromeda galaxy for all stars that have a metallicity of [Fe/H] $<-1$ ($g_0 < 25.5$).  We assume a distance to Andromeda of $d = 785$ kpc.  We mark the location of stream D, stream C, the two parts of the North West (NW) stream and the Southern Giant (SG) stream. We additionally mark the locations of three known dwarfs (N185, N147, and M33) as well as Cloud E and the SW Cloud. The dotted ellipses show two examples of the region sizes we use in this paper. The dashed ellipse has an angular radius of $r_{\rm angular} = 1.825 \deg$, which corresponds to 25 kpc at the distance of M31.  This is one of the region sizes we use to search for known dwarf debris. The solid ellipse has $r_{\rm angular} = 0.365 \deg$, which corresponds to 5 kpc at the distance of M31. Note that both ellipses are circles in spherical sky coordinates, but appear ``squashed" here due to the high declination. 
}
\label{fig:m31}
\end{figure}

\section{Data}\label{sec:data}
PAndAS is a photometric survey of the stellar disk and halo surrounding our neighboring spiral galaxy, M31 \citep{mcconnachie2009,McConnachie18}. The observations for the survey were carried out using the 1-square-degree field-of-view (FOV) MegaPrime/MegaCam camera on the 3.6m Canada-France-Hawaii Telescope (CFHT) and cover 400 square degrees. PAndAS surveyed in the $g$ and $i$ bands to
depths of $g = 26.5$ mag, $i = 25.5$ mag, with a 50$\%$ completeness in the $g$ and $i$ bands of $\approx$ 24.9 and 23.9, respectively \citep[see Figure 4 in][]{martin16}. Each individual star is resolved with a signal-to-noise ratio of at least 10. We show the PAndAS data \citep[][]{ibata14} in Figure \ref{fig:m31}. \citet{ibata14} divided the data into 406 overlapping regions (see their Fig. 1). In Figure \ref{fig:m31} we have applied a metallicity cut of [Fe/H]$< -1 $. At this cut, the enhancement of stars in the overlapping regions is visible (see 1 $\times$ 1 degree fields). We handle these artifacts in post-processing when we search for linear features in the data.

Several dwarf galaxy stellar streams have been discovered in the stellar halo of M31 \citep[e.g.,][]{ibata07,mcconnachie2009,ibata14}. The most prominent dwarf stream is the Giant Southern Stream first discovered by \citet{ibata01} (see Figure \ref{fig:m31}). Several groups have since identified the B, C, D, and NW streams (see labels on Figure \ref{fig:m31}), which are all likely associated with accreted dwarf galaxies based on the stream metallicities and widths \citep[see e.g.,][]{chapman08, gilbert09}. Near M31, there is also debris emerging from known dwarfs that are in the process of being tidally torn apart by M31's gravitational potential at present day (e.g., M33, NGC 147: \citealt{denja14}).

Throughout the paper, we divide the PAndAS data set into smaller regions. Our region sizes are always at least $10~ \times$ larger than the width of the stream we are searching for, to ensure that our target structures do not fill the region as a large-scale overdensity instead of a stream-like feature. In \citetalias{pearson19}, they injected a synthetic MW Pal 5--like stream to a 10 $\times$ 10 kpc$^2$ PAndAS region, which corresponds to $0.729 \times 0.729~ \deg^2$ at the distance of M31 ($d_{\rm M31}$= 785 kpc).  They updated the number of resolved stars Pal 5 would have at the limiting magnitude of PAndAS ($g_0 < 25.5$). 
Additionally, they scaled the width and length of the stream based on M31's gravitational field and based on the stream's location in M31's stellar halo. Since we can only detect part of the red giant branch (RGB) for Pal 5 at the distance of M31 (see Figure 1 in \citetalias{pearson19}), \citetalias{pearson19} found that a similar stream would be very difficult to detect in the PAndAS data. \citetalias{pearson19} demonstrated, however, that GC streams that are five to ten times more massive than a Pal 5--like stream can be detected in current PAndAS data after a metallicity cut. In this paper, we refer to these synthetic streams as $5{M}_{\rm Pal 5}$ and $10{\rm M}_{\rm Pal 5}$. Motivated by \citetalias{pearson19}, in this paper we use the \citet{astropy13, astropy18} SkyCoordinate module to divide the PAndAS data into equal-area overlapping regions, each with an angular radius of $r_{\rm angular} = 0.729/2 = 0.365~ \deg$ (2766 regions total) when we search for new GC streams. 
Half of each region overlaps with its neighbor region both in the RA and dec directions 
to ensure that linear features at the edge of a region will appear at the center of a neighboring region and not get missed.
To mask high star count objects that are not streams, we use the \citet{martin17}, \citet{McConnachie19} and \citet{huxor14} catalogs to identify known dwarf galaxies and GCs in the PAndAS data. 

\begin{figure*}
\centerline{\includegraphics[width=\textwidth]{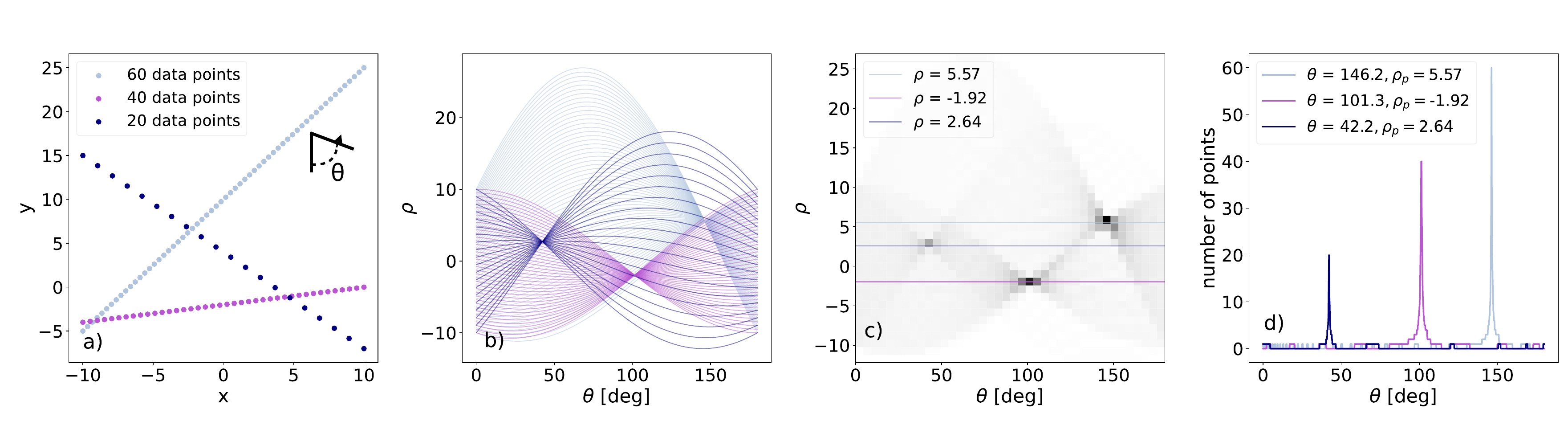}}
 \caption{ Panel (a): three lines with three different orientations made up from 60 (light blue), 40 (purple), and 20 (navy) data points plotted in position space ($x,y$). Panel (b): the Hough transform (see Eq. \ref{eq:HT}) of each point in position space ($x,y$) using an angle, $\theta$ from $0\deg-180\deg$ with 0.1 $\deg$ spacing. Each point in ($x,y$) space is represented by one full sinusoid in ($\rho,\theta$) space (also referred to as the Hough accumulator matrix).  Panel (c): the same ($\theta,\rho$) accumulator matrix as in panel (b), but now binned in $\rho$ to facilitate peak finding. In this example $\Delta\rho = 0.1$ deg, which is the same as the spacing along the $\theta$-axis: $\Delta\theta$. The horizontal lines indicate the $\rho$-value at which each of the three lines have the maximal amount of overlapping sinusoids in panel b. Panel (d): the value of the accumulator matrix along each of the horizontal lines in panel (c) as a function of angle. The accumulator peaks occur at $\theta = 50.8 \deg$ (navy line), $\theta = 101.3 \deg$ (purple line), and $\theta = 146.2 \deg$ (light blue line). The maximum value of each line corresponds to the number of points making up each line in panel (a).
}
\label{fig:input}
\end{figure*}

\section{The \texttt{Hough Stream Spotter}}\label{sec:RHT}
\citet{clark14} developed the \texttt{Rolling Hough transform} (\texttt{RHT}) machine vision algorithm, which quantifies linear structure in 2D image data. They applied the \texttt{RHT} to measure the orientation of filamentary structure in high-resolution’s  Galactic neutral hydrogen emission \citep[see also][]{clark19}. The publicly available \texttt{RHT} code has since been widely used in the astronomical community to quantify linear structure in images of molecular clouds \citep{malinen2016,Panopoulou16}, magnetohydrodynamic simulations \citep{Inoue16}, depolarization canals \citep{jelic2018}, the solar corona \citep{boe2020}, and supernova remnants \citep{raymond2020}, among others. One of the adaptations in the RHT \citep[]{clark14}, as distinct from the classical Hough transform, is to operate on circular subsets of data, rather than on one large rectangular image.
The RHT, working on image-space data, computes a Hough transform for a circular region of sky centered on each pixel in the image, because the goal of the RHT is to parameterize local image-space linearity rather than detect individual lines globally \citep[see][for details]{clark14}. Here our goal is to detect streams -- lines in the distribution of stars that may have a curvature globally -- and so we tile the sky with overlapping circular regions. 
Additionally, rather than operating on pixelated image data, the code presented in this work, the \texttt{HSS}, transforms the individual positions of resolved stars. This allows us to optimize the code for the detection of stellar streams, but the algorithm works on any data that can be described as a list of positions. 

In this Section, we describe the principles of our code (see Sections \ref{sec:HT}, \ref{sec:grid}), and our detection significance and detection criteria (Section \ref{sec:bkg}). 

\subsection{The Hough transform}\label{sec:HT}
The Hough transform \citep{hough62} maps from position space, ($x,y$), to ($\theta,\rho$) space through the following parameterization of a straight line: 

\begin{equation}
    \rho = x~{\rm cos}(\theta) + y~{\rm sin}(\theta)
    \label{eq:HT}
\end{equation}
such that each point ($x,y$) is represented by a sinusoidal curve in ($\theta,\rho$), where $\rho$ represents the minimum Euclidean distance from the origin in ($x,y$) space to the line, and $\theta$ represents the orientation of each possible line in $[0, \pi)$ measured counterclockwise from the vertical.

We illustrate the application of the Hough transform in Figure \ref{fig:input}, where we first plot three different lines in position, ($x,y$), space with three different orientations, made up of 60 points (light blue), 40 points (purple), and 20 points (navy), respectively (see panel (a)). We discretize the set of possible line orientations into an array $\theta_{\rm arr}$ that spans $0$ to $180 \deg$ spaced by $\Delta \theta = 0.1\deg$ and transform each point in Panel a via the Hough transform (Eq. \ref{eq:HT}). In panel (b), each point from panel (a) corresponds to a sinusoidal curve. Thus, the line with 60 data points (light blue) is represented by 60 different sinusioids. For each of the three lines in panel (a), the sinusoidal curves overlap at the same minimal Euclidean distance from the origin, $\rho$, and at the same orientation angle, $\theta$ (panel (b)). Thus, a full straight line in ($x,y$) corresponds to a point in $(\theta,\rho)$ space. $(\theta,\rho)$ space is often referred to as the ``accumulator matrix''. We bin this matrix in $\rho$ to facilitate peak finding in panel (c). Here $\Delta\rho = 0.1$, and there are three clear peaks, which correspond to the overlapping sinusoids for the three different lines. The line with the most points (light blue) has the highest intensity peak. 
We plot three horizontal lines at the Hough accumulator peak $\rho$ values that yield the most overlapping sinusoids ($\rho$ = 5.57, -1.92, and 2.64). In panel (d) we illustrate the intensity of the three peaks by plotting the value of the $(\theta,\rho)$ map for each peak-$\rho$ as a function of $\theta$. We clearly see the excess in intensity in $(\theta,\rho)$ space (i.e., the number of overlapping sinusoids) at three specific angles: $\theta = 50.8 \deg$ (navy line), $\theta =  101.3 \deg$ (purple line), and $\theta = 146.2 \deg$ (light blue line). We can directly read off the number of initial points (60, 40, and 20) that make up each line. Thus, instead of searching for lines in position space, we can simply search for peaks in the accumulator matrix $(\theta,\rho)$ space.

\subsection{Detecting streams in $(\theta,\rho)$ space}\label{sec:grid}
\begin{figure*}
\centerline{\includegraphics[width=\textwidth]{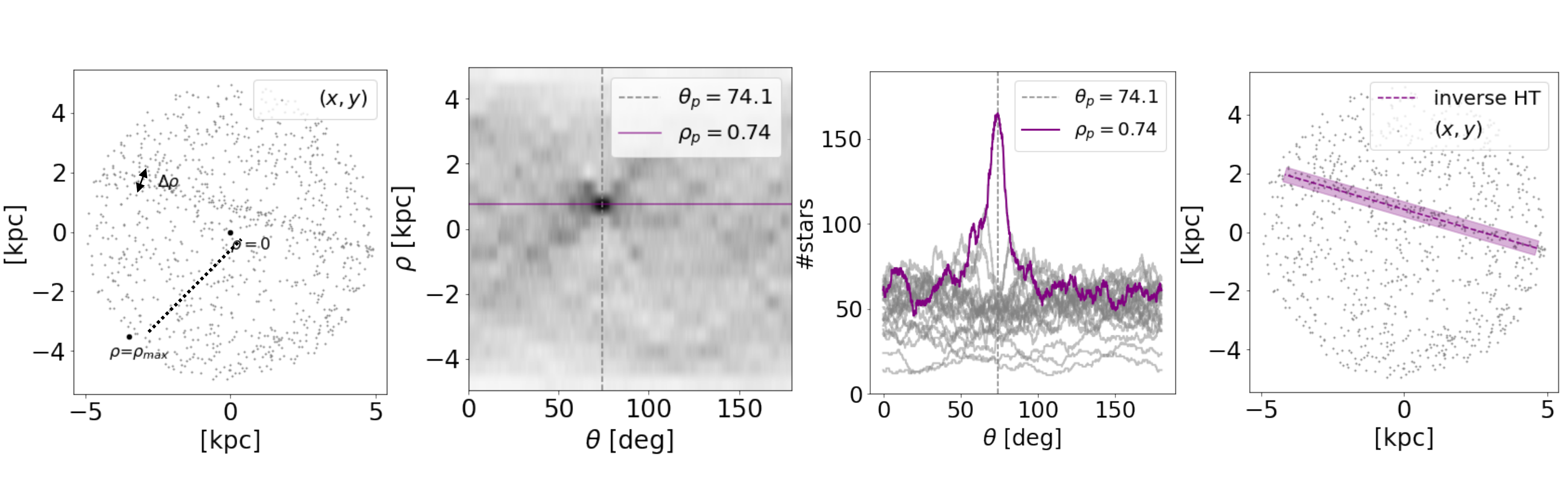}}
\caption{
First panel: a synthetic stream ($N_{\rm *,stream} = 130$) injected to a 5 kpc radius PAndAS region at R$_{\rm gc} = 55$ kpc with $N_{\rm *,background} = 879$ (from \citetalias{pearson19}). $\theta$ is measured counterclockwise from the vertical in ($x,y$) space. $\rho$ is the minimum Euclidean distance from the center, i.e. $\rho_{\rm max} = 5$ kpc is the radius of the region, and $\rho = 0$ is at ($x,y$) = (0,0). $\Delta \rho = 0.5$ kpc is the size of the feature we are searching for, which is here similar to the width of the stream, $w$.
Second panel: the Hough transform (Eq. \ref{eq:HT}) of each star from Figure \ref{fig:input} using an angle $\theta$ from $0-180\deg$ with 0.1 $\deg$ spacing. Each star is represented by a sinusoid. This is similar to the accumulator matrix presented in Figure \ref{fig:input}, but now binned in the $\rho$-direction using $\Delta \rho =0.5$ kpc. The purple dashed line highlights the $\rho$ at which the accumulator matrix has its peak. This corresponds to where in ($\theta, \rho$) space most sinusoids overlap, $\rho_p$. The gray dashed line shows the angle, $\theta_p$, at this same bin. Note the gradient toward higher intensity (darker colors) toward the center (see discussion of this in Section \ref{sec:bkg}). 
Third panel: the value of the accumulator matrix (second panel) as a function of $\theta$ at the peak value of $\rho$ (purple) as well as at all other values of $\rho$ (gray).
Fourth panel: the input region with the recovered inverse Hough transform stripe from ($\theta_p, \rho_p$). This stripe has a width, $\Delta \rho$.
}
\label{fig:demo}
\end{figure*}

To illustrate how the \texttt{HSS} works for stream-finding, we inject a $10{M}_{\rm Pal 5}$ synthetic stream (same age $= 11.5$ Gyr and metallicity [Fe/H] $= -1.3$) from \citetalias{pearson19} to a PAndAS region located 50 kpc from M31's galactic center (Figure \ref{fig:demo}, left). Note that this type of stream should be visible in the PAndAS data by eye if such a stream exists in the stellar halo of M31 (\citetalias{pearson19}). In Section \ref{sec:completeness} we explore the \texttt{HSS}'s ability to detect lower surface density streams and investigate which type of streams the \texttt{HSS} is sensitive to in PAndAS data.
The synthetic stream in this example is ten times more massive than Pal 5, which we take into account when we compute its length, width, and the number of resolved stars in the synthetic stream at the limiting magnitude of PAndAS (see \citetalias{pearson19}, Fig. 1 for details). Due to the 50$\%$ completeness in the $g$ and $i$ bands at  24.9 and 23.9 mag, respectively \citep[see Fig. 4 in][]{martin16}, in this paper, we only use 50\% of the stream stars used in \citetalias{pearson19}. The $10{M}_{\rm Pal 5}$ synthetic stream in \citetalias{pearson19} had 623 resolved stars (see their Fig. 1 upper, right panel). Here, we inject a stream with only 311 stars, and only 130 of these stars fall within the region size used in this example. 
The PAndAS region has an angular radius of  $ r_{\rm angular} = 0.365~ \deg$, which corresponds to a radius of 5 kpc at the distance of M31 (d = 785 kpc). We have applied a metallicity cut of [Fe/H] $< -1$ to this region.

We use Eq. \ref{eq:HT} to Hough transform the positions of stars in our example region. 
The \texttt{HSS} detects streams by finding peaks in the binned ($\theta,\rho$) space (see Figure \ref{fig:input}, panel (c)). Because a stream has a physical width, $w$, the overlap of the sinusoidal curves of its constituent stars will not be a single point in ($\theta,\rho$). We therefore select a scale, $\Delta \rho$, at which we search for linear structures. In this section, we use $\Delta \rho = 0.5$ kpc (see Figure \ref{fig:demo}, left), which is slightly larger than the width of the stream shown in Figure \ref{fig:input} ($w$ = 0.273 kpc in this example from \citetalias{pearson19}). In Section \ref{sec:optimize}, we optimize $\Delta \rho$ for M31 GC stream detection. We show the result of the binned Hough transform in the second panel of Figure \ref{fig:demo}. 
The horizontal and vertical lines highlight the peak in ($\theta,\rho$) space. The value in each bin corresponds to the number of sinusoidal curves, i.e. stars, crossing this particular bin (here darker colors mean more stars). Note that a peak in the $(\theta,\rho)$ grid corresponds to a linear real-space ``stripe'' of width $\Delta \rho$ in $(x, y)$, as illustrated in Figure \ref{fig:demo} (right), where we plot the inverse Hough transform based on the peak ($\theta,\rho$) values. 

A stream-like structure will similarly have an extent in the $\theta$-direction. We refer to this as $\Delta \theta_{\rm smear}$. The minimum number of consecutive bins $\theta$ spans in degrees (see Figure \ref{fig:demo}, second panel) $\Delta \theta_{\rm smear}$, depends on the region size, i.e. $\rho_{\rm max}$ (here 5 kpc), and $\Delta \rho$ (here 0.5 kpc). In the scenario where $\Delta \rho \approx w$ (where $w$ is the width of the linear structure we search for), the minimal $\Delta \theta_{\rm smear} \approx \frac{\Delta \rho}{2\rho_{\rm max}}$. 
In the \texttt{HSS}, we update $\Delta \theta_{\rm smear}$ based on the input region size, $\rho_{\rm max}$, and the spacing in $\rho$, $\Delta \rho$. 

In the third panel of Figure \ref{fig:demo}, we plot the value of the ($\theta,\rho$) map for the Hough accumulator peak $\rho$ value as a function of $\theta$ (purple) as well as for all other values of $\rho$ (gray). This is similar to panel (c) in Figure \ref{fig:input}, except that we here have a background of stars and that the linear feature (the stream) has a physical width. The purple line has a maximum value at ($\theta_p,\rho_p$) = 165 stars. We again see the extent (smear) of the peak in the $\theta$-direction as described above. 
The average value of the purple line off of the peak (here excluding $\theta = 60 - 90$ deg) is $\approx 58$ stars. In physical space, investigating the values of the purple line off of the peak is equivalent to looking at a ``stripe'' with a width of  $\Delta\rho$ at the same minimum Euclidean distance, $\rho$, as the stream, but at a different angle. The purple line therefore includes some of the stream stars, as these will be captured in the ``stripes'' at the off peak angles, which is why the purple line has a higher average value than the gray lines in the third panel. For comparison, the average of all gray lines (i.e., at all other values of $\rho$ than $\rho_p$) is $\approx 50$ stars. We can assess the initial amount of stars that make up the injected stream from the peak value of the purple line. As opposed to Figure \ref{fig:input}, where we did not have a background of points, we here need to take into account the contrast to the background. In the example here, we obtain $\approx 165 - 50 = 115$ stars in the initial input stream, which is similar to the true value of 130 stars. 

Note that the injected stream in Figure \ref{fig:demo} is overdense by more than a factor of three as compared to the background. The \texttt{HSS} can detect streams in M31 with much lower significance. To test this, we injected a stream with Pal 5's width (127 pc), length (12 kpc), and number of stars (34) calculated at a galactocentric radius of 55 kpc in PAndAS \citep[see table 1 and Figure 1 in][]{pearson19}. The \texttt{HSS} successfully flags this stream at the correct ($\theta,\rho$), however if we remove more of the stars, the stream is not distinguishable against the background. Thus, in principle the \texttt{HSS} is sensitive to  Pal 5--like streams in the PAndAS data, but since the significance of the detections is very low, noisy features will also be flagged as streams with this detection threshold. See how we optimize our blind search for GC streams in Section \ref{sec:optimize}, and test the completeness of our method in Section \ref{sec:completeness}.

\subsection{Significance of a stream candidate and criteria to flag the detection}\label{sec:bkg}
To facilitate a blind search for undiscovered GC stream candidates in M31, we need to estimate the significance of each candidate flagged by the \texttt{HSS}. Depending on where in a region the linear structure is located, the area that the stripe can cover will vary (see Figure \ref{fig:demo}). Because the length of the stripe will be shorter toward larger $\rho$ (the edge of the circular region), in Figure \ref{fig:demo} (second panel) we see a gradient of higher bin values toward the center of the $(\theta,\rho)$ grid. Thus, in our assessment of the significance of a stream candidate detection, we need to take the area that each stripe can cover into account at any given $\rho$. For each bin in the $(\theta,\rho)$ grid (which, in the second panel of Figure \ref{fig:demo}, has 36,000 bins), we can express the area, $\delta A$, that the corresponding stripe can cover in ($x,y$) space analytically as:

\begin{equation}\label{eq:dA}
\begin{aligned}
    \delta A = \left( \rho_{\rm max}^2{\rm cos}^{-1}\left(\frac{\rho_1}{\rho_{\rm max}}\right) - \rho_1\sqrt{\rho_{\rm max}^2 - \rho_1^2}\right)\\
    -  \left( \rho_{max}^2{\rm cos}^{-1}\left(\frac{\rho_2}{\rho_{\rm max}}\right) - \rho_2\sqrt{\rho_{\rm max}^2 - \rho_2^2}\right)
\end{aligned}
\end{equation}
where 
$\rho_1 = \rho - \frac{\Delta \rho}{2}$,  $  \rho_2 =  \rho + \frac{\Delta \rho}{2}$, 
$\rho_{max}$ is the radius of the region, and $\Delta \rho$ is the bin size in $\rho$. $\delta A$ is independent of $\theta$ due to rotational symmetry. 

Any given region contains a total number of stars, $N_{\rm stars}$. To assess the significance of a detection in $(\theta,\rho)$ space, we need to ask: what is the probability that $k$ or more stars could fall in a given bin by chance? Here, $k$ is the actual number of sinusoids (i.e. stars) crossing a given bin in $(\theta,\rho)$ space (i.e. the value in the individual bins in Figure \ref{fig:demo} second panel). The probability of there being $k$ stars in certain bin, $i$, is related to the area that a certain stripe covers in ($x,y$) space under the assumption that the region is well represented by a uniform field of stars. Thus, the probability of there being $k$ stars in an area $\delta A$ (and $N_{\rm stars}-k$ stars in the rest of the region) can be expressed as the probability mass function of the binomial formula, with $p =(\delta A/A)$:

\begin{equation}
\begin{aligned}
    Pr (X = k) = \binom{N_{\rm stars}}{k} p^k (1-p)^{N_{\rm stars}-k}
\end{aligned}
\end{equation}
where the maximal number of stars that can fall in any given bin is the total number of stars in the region, $N_{\rm stars}$. 

We can then ask: what is the probability that $k$ or more stars should fall in a certain bin by chance? For each given bin,  this can be expressed as:
\begin{equation}
\begin{aligned}\label{eq:binomial}
    Pr (X \geq k) = \sum_{i = k}^{i = N_{\rm stars} }\binom{N_{\rm stars}}{i} p^i (1-p)^{N_{\rm stars}-i}
\end{aligned}
\end{equation}
Thus, given the values, $k$, in each bin of the $(\theta,\rho)$ grid for the data (see Figure \ref{fig:demo}), the total number of stars in a region, $N_{\rm stars}$, and the area that each stripe covers, $p = \delta A/A$, we can compute a $(\theta,\rho)$ grid of the probability for each bin having the value $k$ or more stars in each data bin. If the probability that a certain bin has $k$ or more stars by chance is very low, we flag this bin as a possible detection. In the limit of a large number of stars, $N_{\rm stars}$, Equation \ref{eq:binomial} approaches a Poisson distribution. Because we apply the \texttt{HSS} to subregions, global gradients and large substructures in halos are negligible, and the assumption of a uniform distribution of the background stars is valid. There are many choices that can be made in terms of handling the background. An alternative approach to assess the significance of the peaks in $\theta,\rho$ space is to compare the value of the peak to the surrounding values in $\theta,\rho$ space \citep[see, e.g., Figure 7 in][]{shih21}.

\begin{figure}
\centerline{\includegraphics[width=\columnwidth]{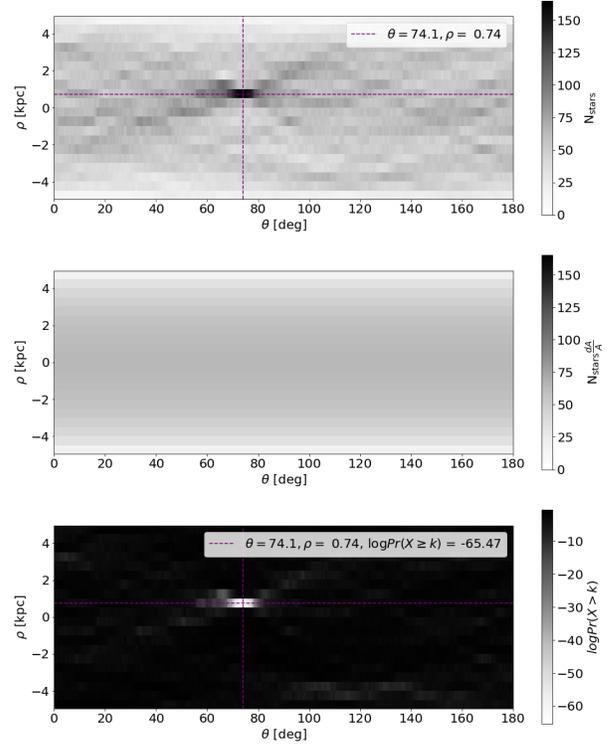}}
\caption{Top panel: same $(\theta,\rho)$-grid as in Figure \ref{fig:demo} based on the Hough transform (Eq. \ref{eq:HT}) of all $N_{\rm stars} = 1009$ total stars in Figure \ref{fig:input} (left), where the gray scale shows the number of stars, $k$, crossing each of the 36,000 bins.
Middle: the number of stars that should fall in each bin if the stars are distributed uniformly in the region with the probability $p=\frac{dA}{A}$ (see Eq. \ref{eq:dA}).
Bottom: the probability of the ($\theta,\rho$) data grid (top) having $k$, or more stars crossing each specific bin, by chance (Eq. \ref{eq:binomial}). The purple dashed lines highlight the flagged stream detection (corresponding to the stripe in Figure \ref{fig:demo}, right). This bin had probability $< 3.4 \times 10^{-66}$ of having $k$ stars (see color bars) in that data specific bin, by chance, and is the only significant outlier in this probability distribution.}
\label{fig:dA_demo}
\end{figure}

In Figure \ref{fig:dA_demo}, we again show the $(\theta,\rho)$ grid for the synthetic stream injected to PAndAS data (top, which is the same as the second panel of Figure \ref{fig:demo}), as well as the $(\theta,\rho)$ grid for $N_{\rm stars} \frac{\delta A}{A}$ from Equation \ref{eq:dA} (middle), and log$_{10}$ of the binomial probability distribution from Equation \ref{eq:binomial} based on the two top panels (bottom). Note that we use log$_{10}$ of the probability to avoid machine precision errors. In this example, the flagged synthetic stream detection is in the bin where $\theta = 74.1\deg$ and $\rho = 0.74$ [kpc] (see purple dashed lines), and has log$_{10}$Pr$( X \geq k) =  -65.47$, i.e. the probability of the data showing the value $k$ or higher in that specific bin, by chance, was  $<  10^{-65.47}$. Thus, instead of searching for peaks in the ``number of stars'' ($\theta,\rho$) space (e.g., as presented in panels two and three of Figure \ref{fig:demo}), we can instead search for peaks in the binomial probability ($\theta,\rho$) space, where we have already taken into count the area that a stream can have and the total number of stars in the background.

Motivated by our intuition from the synthetic stream in the above example, we use the following criteria to flag a stream candidate with the \texttt{HSS} code:
\begin{itemize}
    \item {\it Significance}: a probability threshold in the binomial distribution as described in Eq. \ref{eq:binomial} defined as log$_{10}$Pr$( X \geq k) <$ Pr-thresh.
    
    \item {\it Size}: the detection must span at least  $\Delta \theta_{\rm smear}=\frac{\Delta \rho}{2 \rho_{\rm max}} \deg$ in $\theta$ in the $(\theta,\rho)$ grid (see Figure \ref{fig:dA_demo} top panel).
    
    \item {\it Uniqueness}: a $\theta$-separation of peaks by at least $10~ \deg$ in $\theta$, so that we do not flag the same linear structure multiple times.
    
    \item {\it Overlap}: an edge criterion of $\rho < \rho_{\rm max} - \rho_{\rm edge}$, where $\rho_{\rm max}$ is the size of the region, to avoid flagging overdense features at the edges of regions. 
\end{itemize}

For each region, \texttt{HSS} saves a figure of the input data region and a figure of the input data with any flagged stream detections (as in Figure \ref{fig:demo} right panel). The code additionally stores the binomial probability distribution of bins (see \ref{fig:dA_demo} lower panel).
If there is a stream detection, \texttt{HSS} stores the plots starting with the filename \texttt{Stream}. If there are more than ten flagged stream detections in one region, this means that we have likely detected a ``blob'', as spherical objects will have overdensities in ($\theta,\rho$) space along a sinusoid covering all angles. For these cases, we name the files \texttt{Blob} and do not count them as a stream candidate detection. If there are consistently 10 or more flagged streams in each region, that can also indicate that our Pr-thresh value is too high (such that we find multiple peaks that are actually noise). If there is no detection, the \texttt{HSS} outputs a filename called  \texttt{Empty}.

\section{Application to PAndAS data}\label{sec:applypandas}

\subsection{Input data}\label{sec:input}
Before we feed our data regions (see Section \ref{sec:data}) into \texttt{HSS}, we transform to spherical sky coordinates $(X,Y)$:

\begin{equation}
\begin{aligned}\label{eq:sphere}
    X = 
    \frac{\rm cos(\delta) * sin(\alpha - \alpha_0)}{\rm cos(\delta_0)*cos(\delta)*cos(\alpha-\alpha_0) + sin(\delta_0)*sin(\delta)}
\end{aligned}
\end{equation}

\begin{equation}
Y = \frac{\rm cos(\delta_0) * sin(\delta) - cos(\delta) * sin(\delta_0) *cos(\alpha-\alpha_0)}{\rm sin(\delta_0)*sin(\delta) + cos(\delta_0)*cos(\delta)*cos(\alpha - \alpha_0)}
\end{equation}
where $\alpha = $ R.A., $\delta = $ decl., and $\alpha_0$, $\delta_0 $ are the tangent points of each region projection (i.e. the center of whatever region you are projecting). This means that each data region that we input to \texttt{HSS} will be a circle with $r_{\rm angular} = 0.365~ \deg$ with an origin at $(X,Y) = (0,0)$. All regions are spaced uniformly on the surface of the R.A./decl. sphere, which ensures both equal areas of all regions, and that \texttt{HSS} does not preferentially detect linear structure in one spatial direction.
\texttt{HSS} allows the option to use sky coordinates and read in data sets in degrees, or the code can work with any input unit  and will then ignore sky coordinate transformations.

We mask out dwarf galaxies and GCs in the data, since we are not interested in re-finding known objects. We use {\it Astropy} \citep{astropy13, astropy18} to remove an area of 5 $\times~ r_h$ surrounding each dwarf and GC position \citep[][]{huxor14,martin17,McConnachie19}. In the \texttt{HSS}, there is an option to include your own mask position and size files. 

For regions that intersect with a mask, the ``stripes'' (see Figure \ref{fig:demo}, right) can fall partially within a mask and partially outside a mask. We therefore compute $\frac{\delta A}{A}$ numerically, since this breaks the assumed rotational symmetry in our analytic expression (see equation \ref{eq:dA}). In the numerical case, we populate regions containing masks uniformly with stars, such that each region has at least 100 stars/kpc$^2$. All of these stars are distributed outside of the masks. We then Hough transform each of these stars via Equation \ref{eq:HT}, compute a $(\theta,\rho)$ grid with the same $\Delta \rho$ spacing as the data, and divide by the total number of stars. The value in each bin, $i$, is thus $\frac{\left[n_{\rm random}\right]_i}{N_{\rm stars, total}}$, where $n_{\rm random}$ is the amount of stars that fell in one bin, $i$, and $N_{\rm stars, total}$ is the number of stars in the region. We require a number density of at least 100 stars/kpc$^2$ to ensure a uniform distribution of stars for the numerical dA calculation. The fraction, $\frac{\left[n_{\rm random}\right]_i}{N_{\rm stars, total}}$, is equal to $\frac{\left[dA\right]_i}{A}$, where $A$ is the area of that region. Thus, we now have a numerical representation of $p = \frac{dA}{A}$ for each bin and can use this to calculate the probability in Equation \ref{eq:binomial} and produce a map equivalent to the bottom panel of Figure \ref{fig:dA_demo}. 

\subsection{Optimizing \texttt{HSS} parameters for GC streams in M31}\label{sec:optimize}
In order to optimize \texttt{HSS} to find GC streams in M31, we investigate which $\Delta \rho$ yields the most significant detection in the binomial probability space (see the lower panel of Figure \ref{fig:dA_demo} for the $10{M}_{\rm Pal 5}$ synthetic stream from \citetalias{pearson19}). If there is a low probability of Pr$(X \geq k)$ for a certain bin by chance, this means that we have detected a linear overdensity (see white flagged bins in Figure \ref{fig:dA_demo}, lower panel). 
In this Section, we search for the bin size, $\Delta \rho$, which yields the lowest probability Pr$(X \geq k)$ in the $(\theta,\rho)$ grid for the $10{M}_{\rm Pal 5}$ synthetic stream. Thus, we effectively change the stripe width (see Figure \ref{fig:input}) to determine which width optimizes the detectability of $10{M}_{\rm Pal 5}$ synthetic streams (see Section \ref{sec:num}).
Additionally, we investigate which Pr$(X \geq k)$-threshold to apply to our search in order to detect potential GC streams in the PAndAS data without adding too much noise (see Section \ref{sec:prthresh}).  Note that the stream width in this example is $w = $ 0.273 kpc (see Table 1 in \citetalias{pearson19}). In this Section, we approach these questions numerically. See Appendix \ref{sec:analytic} for an analytic approach with a subset of different stream widths and backgrounds.
 
\subsubsection{Investigating the HSS search width}\label{sec:num}
\begin{figure}
\centerline{\includegraphics[width=\columnwidth]{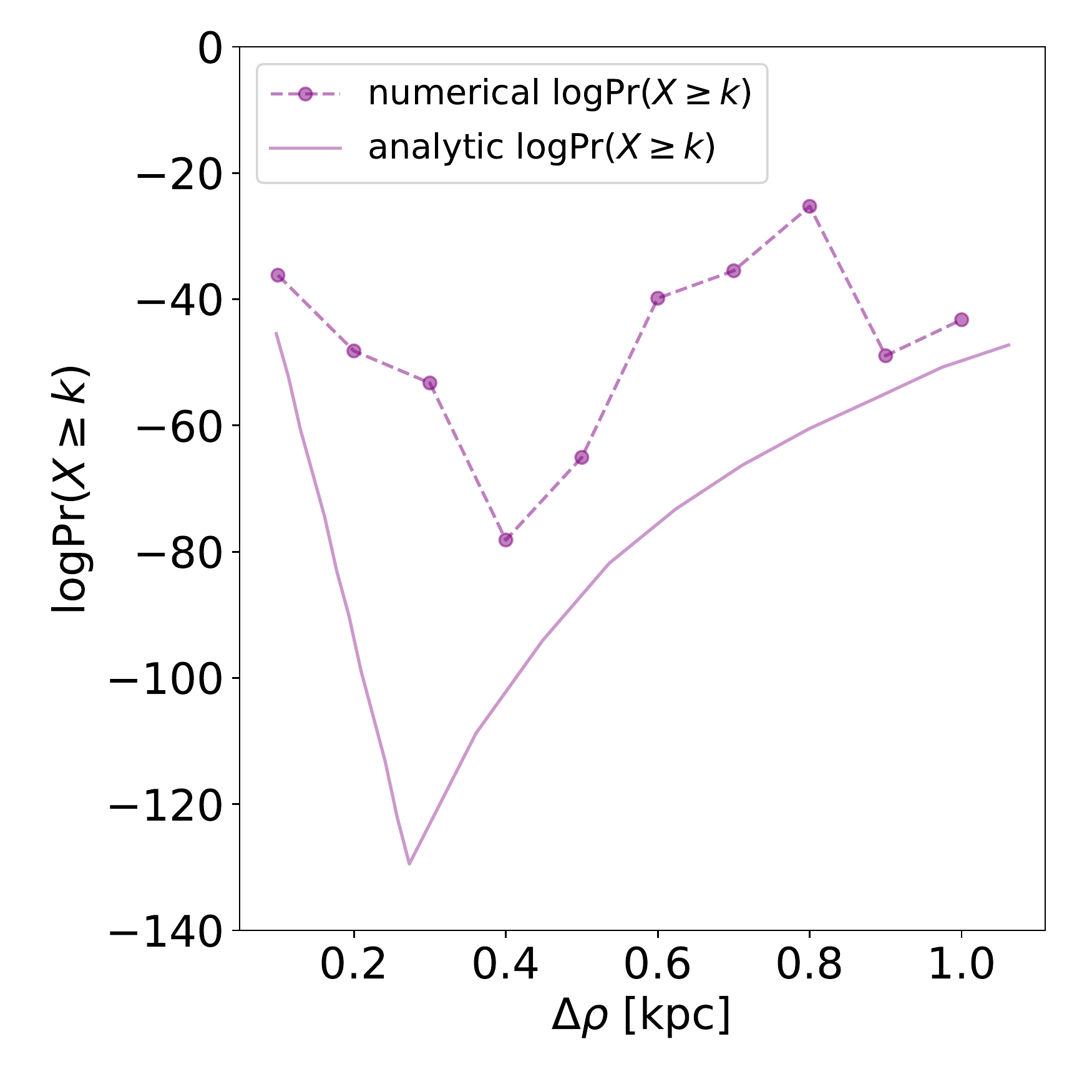}}
\caption{The minimum log$_{10}$Pr$(X \geq k)$ bin value (purple dots) for the $10{M}_{\rm Pal 5}$ synthetic stream in PAndAS data (see Figure \ref{fig:input}) as a function of $\Delta \rho$ in steps of 0.1 kpc. If the probability is very low for a bin having Pr$(X \geq k)$ that means we potentially have a stream detection. For this $10{M}_{\rm Pal 5}$ synthetic stream example, a $\Delta \rho$= 0.4 kpc yields the most significant detection (log$_{10}$Pr $\approx -78$), thus we use $\Delta \rho$ = 0.4 when we blindly search for GC streams in PAndAS data. The  magenta solid line shows the analytic counterpart to this example, which has a very similar shape and minimum log$_{10}$Pr-value to the numerical example, but has its minimum at the exact width of the stream (see Appendix \ref{sec:analytic} for details).}
\label{fig:drho}
\end{figure}

We run the \texttt{HSS} on the input data shown in Figure \ref{fig:input} (left) using $\Delta \rho = 0.1-1$ kpc in steps of 0.1 kpc, and show the results in Figure \ref{fig:drho} (magenta dashed line). $\Delta \rho$ = 0.4 kpc yields the most significant detection (i.e. lowest log$_{10}$Pr$(X \geq k)$-value). If $\Delta \rho < 0.4$ kpc, the minimum log$_{10}$Pr$(X \geq k)$-values are all larger than $-55$, and the minimum Pr$(X \geq k)$-values are larger than $-50$ if $\Delta \rho > 0.5$. Note that for the detections with $\Delta \rho < 0.4$ kpc, multiple streams were flagged on top of the actual synthetic stream, instead of one clear peak, as in the case for $\Delta \rho > 0.4$ kpc. This is because  $\Delta \rho$ becomes smaller than the actual width of the stream, so the peaks span several bins in $\rho$ and multiple structures are flagged with slightly different Euclidean distances, $\rho$, from the origin. 

The fluctuation at large $\Delta \rho$ (see Figure \ref{fig:drho}, dashed magenta line) is due to the fact that the synthetic stream can be partially detected in a stripe (rather than covering that whole stripe) depending on the stream's location in the region. To summarize, using a $\Delta \rho$ slightly larger (0.4 kpc) than the width of the stream (0.273 kpc) maximizes the signal from the stream in this example.  The magenta solid line in Figure \ref{fig:drho} shows the analytic version of this line in an idealized case, where the stream is assumed to cross the center of the region to avoid partial overlap between the stream and the stripe (see details in Appendix \ref{sec:analytic}). In this case, the optimal stripe width $\Delta \rho$ is equal to the exact width of the stream. Note how the shape of the lines is very similar between the analytic and numerical examples, but the detection is more significant (lower log$_{10}$Pr-value) for the analytic case, where the stream is assumed to cross the center of the region. In that scenario, the signal is not smeared between several $\rho$-bins. We conclude that a search $\Delta \rho$ width about one to two times larger than the target stream width is optimal to ensure that the stream width is thinner than the stripe (Figure \ref{fig:drho}). This is a user-specified input to the \texttt{HSS}.

\subsubsection{Choosing the HSS Pr-thresh value}\label{sec:prthresh}
Motivated by the fact that $\Delta \rho$ = 0.4 kpc optimizes the stream detection in our numerical example, we use this $\Delta \rho$ to test which Pr$(X \geq k)$-threshold to use to flag the synthetic stream. If we use a high Pr$(X \geq k)$-threshold, we might flag noise in the field as detections, but if we are too conservative and use a very low Pr$(X \geq k)$-threshold, we might miss the stream. To carry out this test, we run \texttt{HSS} with $\Delta \rho$ = 0.4 kpc, and vary the log$_{10}$Pr$(X \geq k)$-threshold (Pr-thresh) from -$120$ to $0$ in steps of $5$. We find that for $-5 \leq$ Pr-thresh $< 0$, \texttt{HSS} detects $>10$ noisy features as well as the synthetic stream in the input field. For $-15 \leq$ Pr-thresh $\leq -10$, the stream is detected at the correct orientation along with two different stream orientations (for Pr-thresh $=-10$) and with one other stream orientation (for Pr-thresh $= -15$). In Figure \ref{fig:dA_demo} (bottom panel), similar lower-significance peaks also surround the minimum. 
For $-65 <$ Pr-thresh $ \leq -20$, we detect just the synthetic stream at the same bin (i.e. same $(\theta,\rho)$ value). For  Pr-thresh $ < -65$, we do not detect the synthetic stream. Note that for this stream, the minimum Pr-thresh $=-78.15$ (see Figure \ref{fig:drho}), but since we have a $\Delta \theta_{\rm smear}$ criteria, the stream needs to be detected above the Pr-thresh in several consecutive $\theta$ bins (see Section \ref{sec:bkg}), which is only the case when Pr-thresh $ > -65$.
Note that this test is based on one field in PAndAS, and that the backgrounds will vary from region to region (in Section \ref{sec:completeness} we test various locations). Since we will run \texttt{HSS} blindly on PAndAS data, using a nonconservative value (e.g., Pr-thresh $= -15$) will allow us to find streams in noisy fields, without flagging too many spurious structures.

From our analytic investigation in Appendix \ref{sec:analytic}, we additionally found that: (1) a higher number of total stars will lead to a higher significance detection, even with a fixed number density contrast between the stream stars and backgrounds stars, (2) a larger contrast between the stream and background yields a large difference in detection significance, and (3) wider streams yield a more significant detection for a fixed number density of stars in the streams. Related to point 3, due to the presence of dark matter in dwarf galaxies, we cannot scale directly from stellar stream densities in GC streams to stellar stream densities in dwarf streams. However, with access to dwarf stream data, we can measure the dwarf streams' stellar number densities, and use Equation \ref{eq:binomial} to calculate which Pr-thresh to apply.

\subsection{Application to PAndAS dwarf streams}\label{sec:dwarfs}
\begin{figure*}
\centerline{\includegraphics[width=\textwidth]{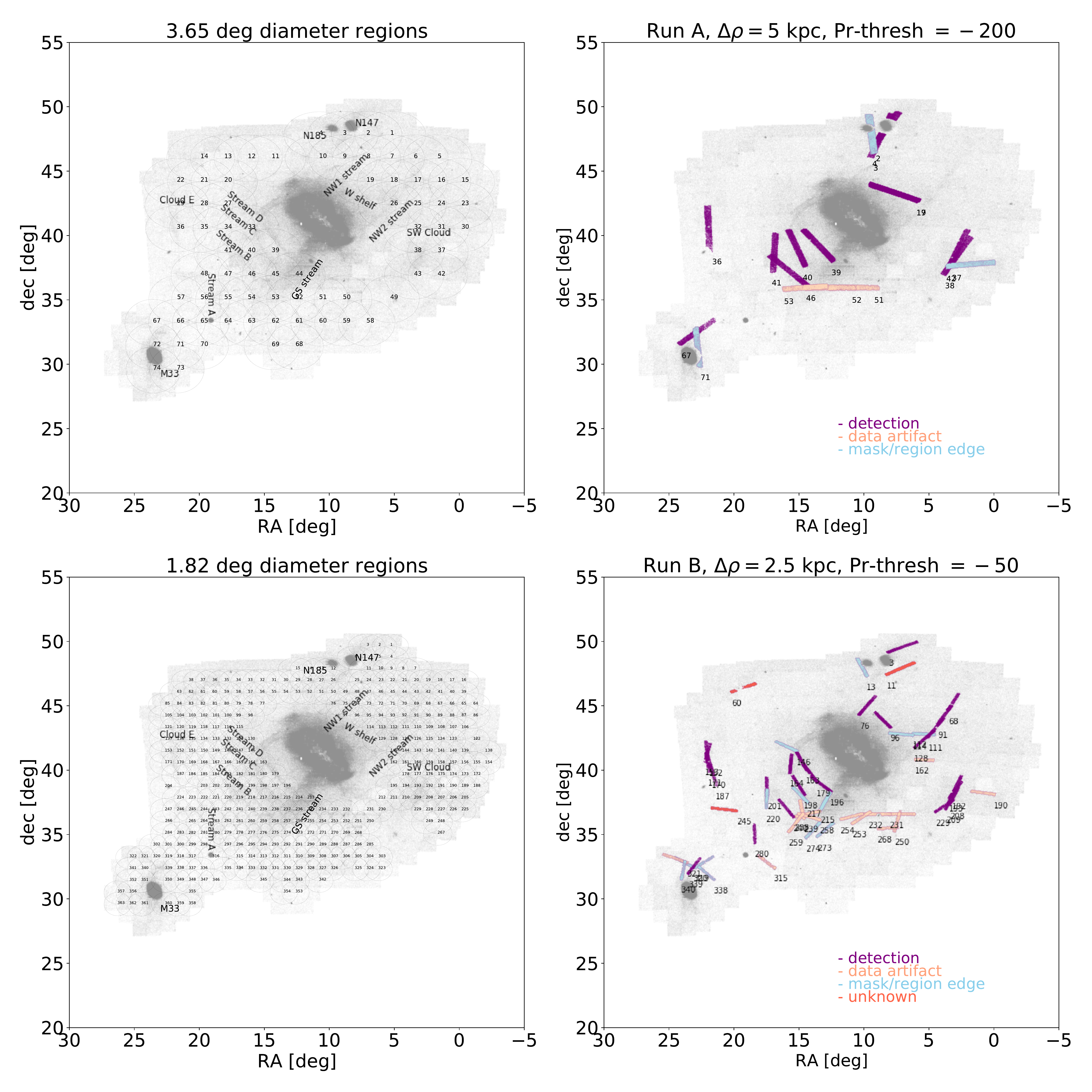}}
\caption{Summary of \texttt{RunA \& RunB}.
Left panels: PAndAS data with [Fe/H]$< -1$ with all known objects labeled (see Figure 12 in \citealt{McConnachie18}), which \texttt{RunA} and \texttt{RunB} should recover. The region sizes are overplotted as ellipses (top: 3.65 deg (${\approx}50$ kpc) in diameter, bottom: 1.82 deg (${\approx}$25 kpc) in diameter). 
Upper right panel: \texttt{RunA} results (region size = 50 kpc, $\Delta \rho$ = 5 kpc, Pr-thresh $< -200$). Purple indicates a known feature detection. All features except for Stream A, the GS stream and the NW stream are recovered in \texttt{RunA}. Pink streams represent data artifacts - linear features at the edges of the CFHT pointings.
Light blue streams show edge detections  - where a feature at the wrong angle is detected, because the feature is close to the edge of that specific region or mask in a region.
Lower right panel:
\texttt{RunB} results (region size = 25 kpc, $\Delta \rho$  = 2.5 kpc, Pr-thresh  $<-50$). Note how the narrow features (e.g., NW1, NW2 streams) are recovered here.  The red streams here show previously undetected features, which we explore in Figure \ref{fig:newdwarfstreams}. Note that we here did not include detections that were labeled as ``blobs" by the \texttt{HSS}, which occurs when more than 10 features are discovered above the Pr-thresh in one region (this was the case for the lower part of the M33 stream and also for the GS stream, due to the location and size of the regions). 
}.
\label{fig:runabsummary}
\end{figure*}

Before we run the \texttt{HSS} blindly on PAndAS data to search for unknown GC stream candidates (see Section \ref{sec:results}), we test whether our code can recover the known, wider debris features in M31 \citep[see structures A through M in][Figure 12, where we omit their NE shelf, E shelf, and G1 clump as these are contained in our M31 mask]{McConnachie18}. In Table \ref{tab:dwarfs} (left), we list the features that we are attempting to recover. We also label these in Figure \ref{fig:runabsummary} (left), where the regions and data are plotted after a metallicity cut of [Fe/H]$< -1$. 

Most of the dwarf streams in M31's stellar halo are a few kiloparsecs wide, with the exception of the GS stream which is ${\approx}0.5\deg$ wide corresponding to ${\approx}6.9$ kpc \citep{mc03}. To capture the range of wide debris features, we divide the PAndAS data set into regions with two different angular extents: first 73 regions with $r_{\rm ang} = 1.825\deg$ (25 kpc at M31's distance; see Figure \ref{fig:runabsummary}, top left)  and second: 358 regions with $r_{\rm ang} = 0.9125 \deg$ (12.5 kpc at M31's distance; see Figure \ref{fig:runabsummary}, bottom left). Most of these regions have neighbor regions that overlap by 50\% in both the R.A. and decl. directions. For regions on the edge of the data sets or on the edge of a large mask (e.g., the M31 mask), there are not overlapping regions in the direction of the edge or the mask (see Figure \ref{fig:runabsummary}, left column).
We transform each region to spherical sky coordinates (see Eq. \ref{eq:sphere}) and run \texttt{HSS} with two different set of parameters (see the definition of each parameter in Section \ref{sec:bkg}):
\begin{itemize}
\item \texttt{RunA}: Region diameter = 50 kpc,
$\Delta \rho = 5$ kpc, Pr-thresh $< -200$,
 $\theta_{\rm smear} = 5.73\deg$, $\theta_{\rm separation} = 10\deg$, $\rho_{\rm edge} = 10$ kpc.
\item \texttt{RunB}: Region diameter = 25 kpc, $\Delta \rho = 2.5$ kpc, Pr-thresh $< -50$.
 $\theta_{\rm smear} = 2.86\deg$, $\theta_{\rm separation} = 10\deg$, $\rho_{\rm edge} = 5$ kpc.
\end{itemize}
We used the NW2 stream (see Figure \ref{fig:runabsummary}) as an example dwarf stream to motivate the difference in the  Pr-tresh-values for \texttt{RunA} (Pr-thresh $< -200$) and \texttt{RunB} (Pr-thresh $< -50$). For the NW2 stream, the number density in the stream is ${\approx}21.5$  stars/kpc$^2$ and the number density in its close vicinity is ${\approx}18.5$ stars/kpc$^2$. The region sizes in \texttt{RunA} and \texttt{RunB} are factors of 25 and 6.5 times larger, respectively, than the region size used in Section \ref{sec:analytic} ($r = 5$ kpc), respectively, and the search widths for the streams, $\Delta\rho$, are $\approx$ 10 and 5 times wider. Thus, we can calculate $p=dA/A$, by scaling up the difference in the areas of the streams and regions. With $p$, the stellar number densities (and thus number of stars) in the NW2 stream, and the stellar number densities in the background in hand, we can use Eq. \ref{eq:binomial} to calculate the analytic minimum log$_{10}$Pr-values for \texttt{RunA} and \texttt{RunB} (see also Appendix \ref{sec:analytic}). We find that the minimum log$_{10}$Pr-values are $-679$ for \texttt{RunA} and $-172$ for \texttt{RunB}. Thus, a factor of four difference for the two runs. For the $10{M}_{\rm Pal 5}$ synthetic stream in Section \ref{sec:optimize}, we found a numerical log$_{10}$Pr-minimum at $-78$ (see Figure \ref{fig:drho}), but showed that Pr-thresh $< -15$ was the ideal threshold to use to detect the $10{M}_{\rm Pal 5}$ synthetic stream without much noise (see Section \ref{sec:optimize}). By comparison, the analytic minimum for this $10{M}_{\rm Pal 5}$ synthetic stream example was $-129$ (see Figure \ref{fig:drho}). Thus, for the two dwarf runs (\texttt{RunA} and \texttt{RunB}), we fix the ratio of the Pr-thresh to 4, but use Pr-thresh $< -200$ for \texttt{RunA} and  Pr-thresh $< -50$ for \texttt{RunB} to ensure that we capture fainter features than the NW2 stream.

For each region with a flagged detection, we only plot the most prominent detection in that region, i.e. the detection with the minimum log$_{10}$Pr-value. In some flagged regions, \texttt{HSS} detects several different features; however, in all flagged regions where a known debris structure was present, that structure was the most  prominent log$_{10}$Pr peak. Thus, if we had used a lower, more conservative Pr-thresh, we would pick up the specific structure only. Note that if a region has more than 10 flagged detections above the threshold, we classify it as a likely ``blob'' and not a stream, since a ``blob'' would span a full sinusoid in $(\theta,\rho)$-space, and therefore likely have several peaks despite the $\theta_{\rm separation} = 10\deg$ criterion (see Section \ref{sec:bkg}). We do not count these ``blobs'' as detections in this work. 

In the right panels of Figure \ref{fig:runabsummary}, we show PAndAS data and overplot all stars that were a part of an \texttt{HSS} detection stripe based on three criteria: (1) purple streams: detection of known PAndAS feature, (2) pink streams: data artifacts at 0 or 90 $\deg$ due to the $1\deg \times1\deg$ field size of the CFHT pointings, and (3) light blue streams: detection at the edge of an overdense feature in a neighboring region. Each feature is labeled by the number of the region that it was detected in (see left panels).

In \texttt{RunA}, the \texttt{HSS} flags 17 detections regions out of the 73 total regions (see Figure \ref{fig:runabsummary}, upper right). In Table \ref{tab:dwarfs}, we summarize which of the known features (see purple ``stripes'' in Figure \ref{fig:runabsummary}) were recovered and how many regions the features were recovered within. 
Additionally, we list how many data artifacts were flagged (see pink `stripes' in Figure \ref{fig:runabsummary}), as well as
overdense features at the edge of the regions (see blue ``stripes'' in Figure \ref{fig:runabsummary}).

\texttt{RunA} detects all streams except for the NW stream, Stream A, and the GS stream. The first two are likely too narrow for the $\Delta \rho = 5$ kpc criterion (recall how the detection signal is less significant if $\Delta\rho$ is much wider than the width of the stream in Section \ref{sec:num}, Figure \ref{fig:drho} and Appendix \ref{sec:analytic}, Figure \ref{fig:drho_analytic}). The GS stream is wider than our search criteria,  with a high surface density, and it takes up a large area of our region sizes. In several regions, the GS stream was therefore classified as a ``blob''. Had we used larger region sizes and an even lower Pr-thresh, we would have detected this as a stream.  Note also that the southern part of M33's stellar debris is not detected here, as it was flagged as a ``blob''.

\begin{deluxetable*}{lcccccccc}\label{tab:dwarfs}
\tablecaption{Summary of \texttt{HSS} dwarf stream searches.}
\tablecolumns{3}
\tablenum{1}
\tablewidth{0pt}
\tablehead{
\colhead{Debris Feature\tablenotemark{a}} &
\colhead{\texttt{RunA}\tablenotemark{b} } &
\colhead{Reg\tablenotemark{c} } &
\colhead{\texttt{RunB}\tablenotemark{d}  } &
\colhead{Reg\tablenotemark{c}} &
\colhead{Accum. Peak} & 
\colhead{Stripe Dens.} & 
\colhead{$<$Off Stripe Dens.$>$} \\
\colhead{} &
\colhead{} &
\colhead{ } &
\colhead{ } &
\colhead{} &
\colhead{(stars)} &
\colhead{(stars/kpc$^2$)} & 
\colhead{(stars/kpc$^2$)}
}
\startdata
Stream D                    & yes & 1 & yes & 4&7750&31&16.5 \\
Stream C                    & yes &  1 & yes & 2 &5387&21.5&13.4\\
Stream B                    & yes &  2 & yes & 2&3633&14.5 &8.8\\
Stream A                    & no   &  -&  yes & 1&648&10.4&6.2 \\
NW1 Stream\tablenotemark{e}  & no &- & yes &1&1688 &27&14.7\\
NW2 Stream\tablenotemark{e}  & no &-&yes  & 5&1409&22.5&14\\
GS Stream  &no  &-& no & - && & \\
NGC 147's stream             & yes &  2 & yes & 1&5145&20.6&13.4 \\
M33's stream                 & yes &  1 & yes & 1&886&14.2&7.7 \\
Cloud E                     & yes &  1 & yes & 5 &2964&11.8&6.2\\
SW Cloud                    & yes &  2 & yes & 5&4139&16.6&10.4 \\
W shelf                     & yes &  1 & yes & 1&5791&23.1&13.6 \\
Data artifacts              & yes& 3 & yes & 10 &&& \\
Edge detections             & yes &3 & yes  & 11&&&\\
Unknown\tablenotemark{f}  & no &-&  yes &  3&&&
\enddata
\tablenotetext{a}{See features in Figure \ref{fig:m31} and \ref{fig:runabsummary} as well as in \citet{McConnachie18}.}
\tablenotetext{b}{Detected in \texttt{HSS} run with $\Delta \rho$ = 5kpc,  Pr-tresh $<-200$.}
\tablenotetext{c}{Number of regions that this feature is detected in.}
\tablenotetext{d}{Detected in \texttt{HSS} run with $\Delta \rho$ = 2.5 kpc, Pr-tresh $<-50$.}
\tablenotetext{e}{This is the thinnest stream.}
\tablenotetext{f}{Not reported by PAndAS team.}

\end{deluxetable*}

\begin{figure}
\centerline{\includegraphics[width=\columnwidth]{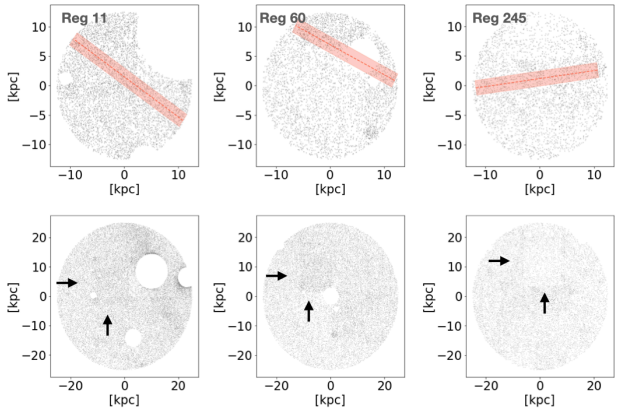}}
\caption{Top panel: the three flagged detections in \texttt{RunB} that are not associated with any known debris features (see red stripes in Figure \ref{fig:runabsummary}). Region 60 (middle) has a flagged detection near a masked out dwarf galaxy (Andromeda XXIV; see white blank circle), which could indicate that we are detecting tidal debris from this dwarf. Bottom panel: we centered the regions on the detected features for region 11 and region 245 and create a 50 kpc diameter region. For region 60, we center the region on the center of the masked out dwarf, Andromeda XXIV, with the same 50 kpc diameter region size. In each case, we see that the flagged features have picked up data artifacts from the size of a CFHT $1\deg \times1 \deg$ pointing, which is apparent in the data due to the incompleteness at the given magnitude and [Fe/H] cut (see black arrows). Thus, these detections are not new dwarf streams, but data artifacts. }
\label{fig:newdwarfstreams}
\end{figure}

\begin{figure}
\centerline{\includegraphics[width=\columnwidth]{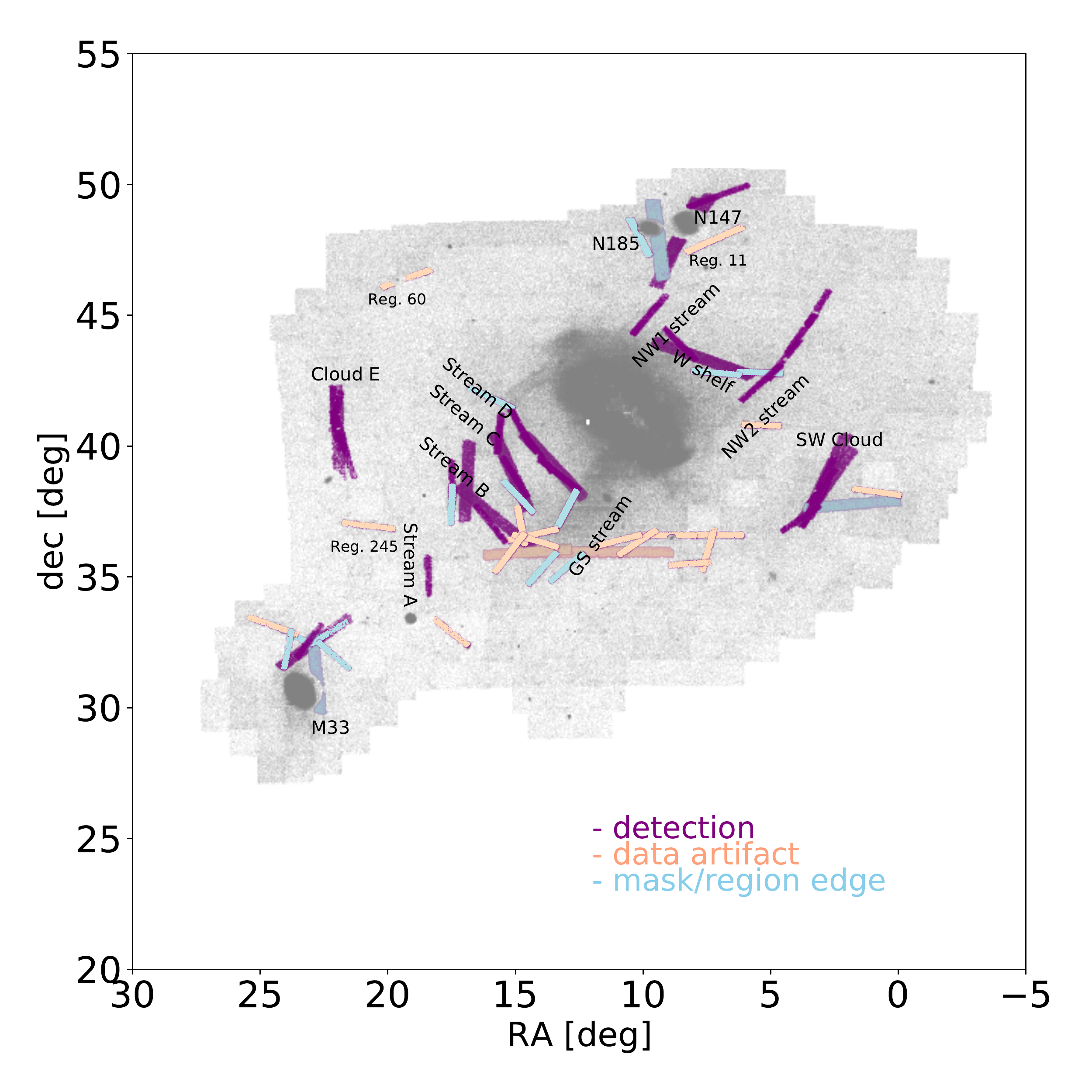}}
\caption{Summary of the findings for \texttt{HSS} \texttt{RunA} and \texttt{RunB}. The wider streams show the results from \texttt{RunA}, and the narrower streams show the results from \texttt{RunB}. The colors are the same as in Figure \ref{fig:runabsummary}. All dwarf streams listed in \citet{McConnachie18} are rediscovered, except for the GS stream due to its wide nature. Regions 11, 60, and 245 have been updated as `data artifacts' as opposed to `unknown' based on the analysis in Figure \ref{fig:newdwarfstreams}. 
Note that using different locations and sizes for the regions would yield slightly different results for the data mask/region edges, as these are detections at the edge of a region. Different Pr-thresh values would also lead to scenarios where the southern part of M33 and the GS stream were labeled as detections instead of ``blobs''.}
\label{fig:dwarfscombined}
\end{figure}

For \texttt{RunB}, where we search for narrower features ($\Delta \rho = 2.5$ kpc), the \texttt{HSS} flags detections in 52 of the 358 regions (Figure \ref{fig:runabsummary} bottom, right).
As for \texttt{RunA}, in Table \ref{tab:dwarfs} we summarize the detection of known features, data artifacts, and
overdense features. We also list whether any detections were flagged where their origin is `unknown' (see red ``stripes'' in Figure \ref{fig:runabsummary}). 
The only stream that the more narrow \texttt{RunB} did not recover was the GS stream, because the stream almost takes up the entire size of the region (see Figure \ref{fig:runabsummary}, lower left) and is therefore flagged as a ``blob'' with more than ten detections. Interestingly, the thinner streams such as NW1 and NW2 are picked up by \texttt{RunB}, which was not the case for \texttt{RunA} (see Figure \ref{fig:runabsummary},  right panels). 

For each detected dwarf feature, we investigate the most significant (minimized log$_{10}$Pr-value) detection of that feature. We investigate the number of stars for each value of $\rho$ as a function of $\theta$ (see purple line in panel three of Figure \ref{fig:demo} as an example) and report the Hough accumulator peak in Table \ref{tab:dwarfs}. This is the number of stars at the values of $\theta_p$ and $\rho_p$ that minimized log$_{10}$Pr (see the peak of purple line in the third panel of Figure \ref{fig:demo} as an example). To assess the density of the detected stream feature, we divide this peak value by the area of the ``stripe'' (250 kpc$^2$ for Run A, and 62.5 kpc$^2$ for Run B). This gives us the stellar density of the dwarf feature. For comparison, we also average over the stellar density for all other stripes positions and orientations, excluding the most significant peak (see all gray lines in Figure \ref{fig:demo} as an example) and list the average value in Table \ref{tab:dwarfs} (last column). We note that the dwarf features are all overdense by more than a factor of 1.5 as compared to their backgrounds at the metallicity cut of [Fe/H] $< -1$.

There were three regions flagged with previously unknown features (see red highlighted stars in Figure \ref{fig:runabsummary}). In Figure \ref{fig:newdwarfstreams}, we show the regions that these stars were detected in (top panels) as well as the flagged stripe detection (red). Particularly region 60 (top, middle) is of interest as it is a flagged detection near a masked out dwarf (Andromeda XXIV), which could potentially could be a dwarf stream not yet reported by the PAndAS team. To analyze these three detections further, we center the regions on each detection (or in the case of region 60, we center the region on the mask center), and create regions with 50 kpc surrounding these new centers (see bottom row). We create these new regions, with the goal of re-running the \texttt{HSS} to check if the streams are flagged again.

However, from Figure \ref{fig:newdwarfstreams} (bottom panels), we notice that each of the detections overlaps with an over- or underdense ``square'' in the PAndAS data (see black arrows). These correspond  to the CFHT $1\deg \times1 \deg$ fields, which are apparent due to the incompleteness of the data at the given magnitudes. The \texttt{HSS} has picked up on the overdensities because they are linear artifacts, and these red detections are not new dwarf streams. When we re-ran the \texttt{HSS} on the new, larger regions (but with the same criteria as for \texttt{RunB}), the code did not flag any detections. Thus, these detections are ``artifacts'', and get picked up when the CFHT $1\deg \times1 \deg$ fields are on the edge of a region (see Appendix \ref{sec:appendix_completeness} where many of these ``squares'' are picked up if we use a Pr-thresh $<-10$).

We combine the results of the two \texttt{HSS} runs in Figure \ref{fig:dwarfscombined}, where the two different width stripe detections are plotted. 
To summarize, the test of the \texttt{HSS} code on known dwarf debris features was successful and rediscovered all known streams and clouds (purple), except for the GS stream, which was flagged as a ``blob'' in both runs due to the high density of stars, the width of the stream and the small region size compared to the stream area. We also report detections of linear artifacts (pink) and  overdense features at the edge of our regions (blue). If  $\Delta \rho$ is narrow, we detect multiple streams on top of the wider tidal features.  Note that when the region size is larger, our assumption that the stars in the field follow a uniform distribution is not entirely correct. With a better estimate of the star distribution or comparisons to stellar halos from simulations \citep[e.g.,][]{bullock05}, we could make a more accurate detection criteria and potentially strengthen our stream signals and detections. However, we leave this for future applications of the code, as our goal in this paper is to find new GC streams. The \texttt{HSS} \texttt{RunA} and \texttt{RunB} did not flag any GC candidates, since we used a very wide $\Delta\rho$. See Section \ref{sec:results} for a narrower search for new GC stream candidates.

\subsection{Completeness checks}\label{sec:completeness}
\begin{figure*}
\centerline{\includegraphics[width=\textwidth]{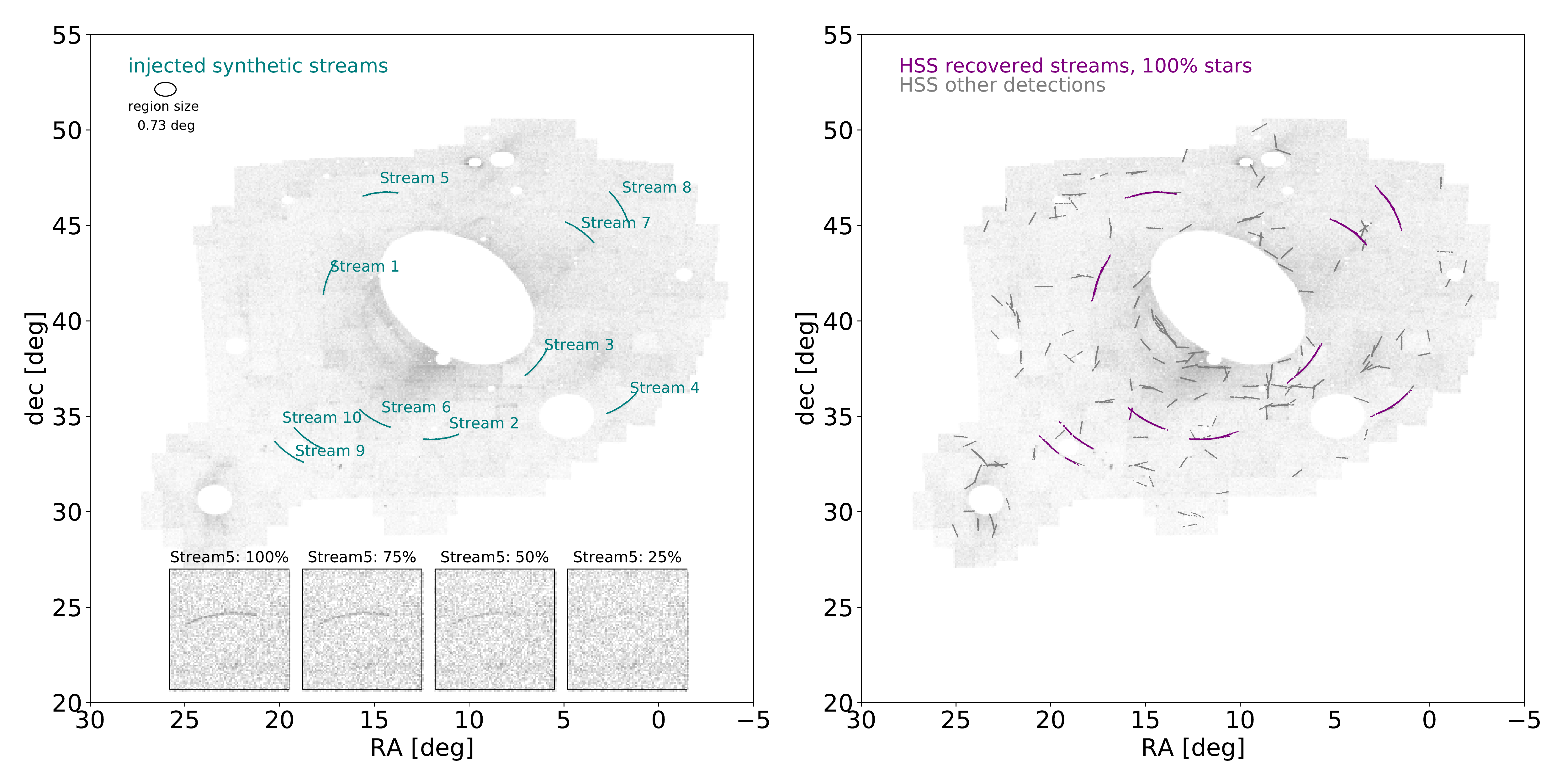}}
\caption{Left: PAndAS data with [Fe/H] $<-1$ (gray), where the M31, dwarf, and GC masks have been removed from the data. We have injected ten $10{M}_{\rm Pal 5}$ synthetic streams at random locations with random orientations (highlighted in teal). The widths of these streams are all 0.273 kpc, the lengths are 25.9 kpc (1.89 deg), and the number of stars in each synthetic stream is 311, motivated from \citetalias{pearson19}. Note that some of the streams are closer to M31 than others, thus residing in a higher-density environment. The four panels in the lower part of the plot show $2\times2\deg$ zooms of Stream1 with 100\%, 75\%, 50\%, and 25\% of its stream stars remaining (from left to right).
Right: The results of a blind \texttt{HSS} search on 0.73 deg overlapping regions (see small ellipse, left) with $\Delta \rho = 0.4$ kpc, Pr-thresh $<-15$  (detections are highlighted in purple). Each of the ten synthetic  streams are recovered by the \texttt{HSS}, and their curvature is captured since the region sizes ($0.73 \deg$ in diameter) are smaller than the length of the synthetic streams ($1.89 \deg$). Thus, should these type of streams exist in the data, the \texttt{HSS} should easily detect them. The dark gray streaks represent other detections by the \texttt{HSS} in this run, which we will discuss in Section \ref{sec:results}.}
\label{fig:completness}
\end{figure*}

In this Section, we investigate the ability of the \texttt{HSS} to detect $10{M}_{\rm Pal 5}$ synthetic streams at various locations in M31's stellar halo, as the density of stars varies across the PAndAS dataset. We inject the $10{M}_{\rm Pal 5}$ synthetic streams with ten random orientations, but ensure that each stream is curved in a concave configuration with respect to the center of M31 to mimic a plausible orbit  \citep[e.g.,][]{johnston01}. The synthetic streams all have a width of 0.273 kpc, a length of 25.9 kpc and initially have 311 stars total (see also Table 1 and Figure 1 in \citetalias{pearson19}), which have been calculated based on the limiting magnitudes of PAndAS at the distance of M31. As mentioned in Section \ref{sec:HT}, we use half of the stars (i.e. 311 instead of 623) presented in Figure 1 in  \citetalias{pearson19}, due to the incompleteness of the PAndAS data \citep[]{martin16}. We test the ability of the \texttt{HSS} to recover the injected synthetic streams if we include a random subset of 100\%, 75\%, 50\%, and 25\% of these 311 stars in the $10{M}_{\rm Pal 5}$ synthetic stream.  

In Figure \ref{fig:completness} (left), we show the location of the ten $10{M}_{\rm Pal 5}$ synthetic streams, each of which consists of 100\% of their 311 stars (teal streaks) injected into the PAndAS data (with [Fe/H] $< -1$). The blank areas represent regions where we have masked out data due to the locations of galaxies or GCs in the PAndas dataset \citep[see][and Section \ref{sec:data}]{huxor14,martin17,McConnachie19}. The four panels in Figure \ref{fig:completness} (left, bottom) show a $2\deg \times2\deg$ zoom of `stream 5' with 100\%, 75\%, 50\%, and 25\% of its stars remaining (left to right). Note that stream 5 is difficult to see by eye if only 25\% of the stars are included. 

We first check whether the \texttt{HSS} can recover these streams with 100\% of the stars in the streams. We divide the data set, which now includes the synthetic  streams, into 2766 overlapping regions with an angular diameter of 0.73 deg (10 kpc at the distance of M31; see small ellipse in Figure \ref{fig:completness} and  details in Section \ref{sec:data}). We then run the \texttt{HSS} with $\Delta \rho = 0.4$ kpc, Pr-thresh $< -15$ motivated from Section \ref{sec:bkg}. We additionally require that $\theta_{\rm smear} > \frac{\Delta \rho}{2\rho_{\rm max}} > 2.86 \deg$, and $\theta$-separation $>10\deg$ (see Section \ref{sec:bkg}). We remove edge detections where $\rho>3$ kpc, as the 2766 regions overlap by 50\% in both R.A. and decl. (see Section \ref{sec:data}),  anything on the edge of one region will be in the center of another region. 

In Figure \ref{fig:completness} (right), we show the results of this run. Here, the purple streaks highlight the stars that have been flagged as part of the detected streams in the blind \texttt{HSS} run. All of the ten synthetic streams are detected by the \texttt{HSS}, and even the curvature of the streams is captured, as the streams span several 0.73 deg regions ( $l_{\rm stream}= 1.89 \deg$), each with a slightly different \texttt{HSS} angle of detection. If M31 indeed has these $10{M}_{\rm Pal 5}$ streams in its stellar halo, the streams will be detected very clearly by the \texttt{HSS}. These streams, however, would also be noticeable by eye (see Figure \ref{fig:completness}, lower left). 

The darker gray, thin streaks in Figure \ref{fig:completness} (right) highlight other \texttt{HSS} flagged detections in this particular run. Note that some of these features appear to be artifacts (at 0 and 90 $\deg$) of the CFHT pointings due to the incompleteness of the data at this metallicity cut, and some features are on top of known dwarf debris features. We will investigate these features in Section \ref{sec:results}. 

\begin{figure}
\centerline{\includegraphics[width=\columnwidth]{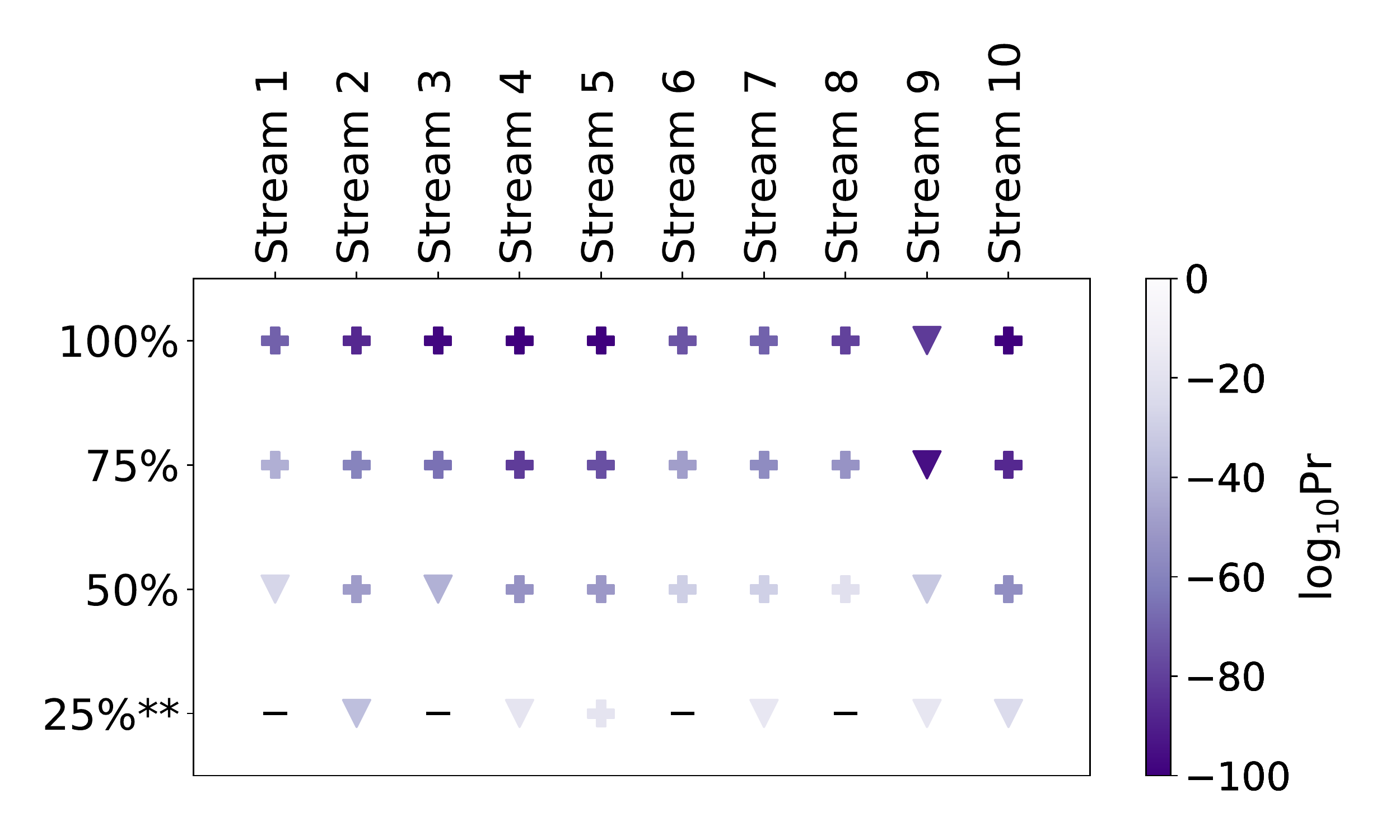}}
\caption{Summary of each of the four different \texttt{HSS} runs with 100\%, 75\%, 50\%, and 25\% of the stars remaining in the ten $10{M}_{\rm Pal 5}$ synthetic streams. For the 100\%, 75\%, 50\% runs, we used a Pr-thresh$<-15$, but for the 25\% run (**), we used a  Pr-thresh $<-10$ (see text for details). 
For each of the runs, we mark whether each specific synthetic stream (streams 1 through 10, see Figure \ref{fig:completness}, left) was recovered. The ``+" markers represent streams that were fully recovered by the \texttt{HSS} across all regions. The triangles represent partially recovered streams. The ``-" markers represent streams that were not recovered. The color bar shows the significance of each detection. 
We note that the synthetic streams with more stars are detected to a higher significance (see darker colors). The trends in the significance of detections persist between all four runs (e.g., the outer halo synthetic stream 9 has a dark color in all four \texttt{HSS} runs). The \texttt{HSS} is complete to streams with 50\% of the surface density of $10 {M}_{\rm Pal 5}$.
}
\label{fig:completness_summary}
\end{figure}
We repeat the exercise above, but now instead inject streams with 75\%, 50\%, and 25\% of the initial 311 stars (we keep the length and width the same in this test). We first inject ten 75\% synthetic  streams into the same locations with the same orientations as shown in  Figure \ref{fig:completness} (left), and read out 2766 new overlapping regions, which now include the ten new synthetic streams with 75\% of the stars. 
We re-run the \texttt{HSS} with the same parameters as above. We then repeat this exercise with ten synthetic streams including only 50\% of the stars (155), and then including only 25\% of the stars (77). For the 75\% case, each synthetic stream is visible by eye (Figure \ref{fig:completness}). For the 50\% case, some of the inner streams are hard to make out by eye, even knowing where they are, and for the 25\% case, we cannot see the synthetic streams by eye.

We summarize the results of all four \texttt{HSS} runs with the ten different synthetic streams in Figure \ref{fig:completness_summary}, where we plot the detection significance of each synthetic stream (1 through 10) for each of the four runs (100\%-25\%). For each \texttt{HSS} run (100\%-25\%) and for each stream (1 through 10), we recorded the one most significant log$_{10}$Pr-value, and the color of each marker represents this significance (see the color bar). 

In the case with ten injected synthetic streams that have 75\% of the stream stars, all ten synthetic streams are recovered across several regions. The streams that are fully recovered are labeled with ``+'' markers in Figure \ref{fig:completness_summary}.  As stream 9 is located close to a masked out dwarf, only part of the stream is recovered (see triangles in Figure \ref{fig:completness_summary}). Note how the significance of each detection is lower in the 75\% run (see fading colors). For the case of the 50\% synthetic streams, all streams were recovered, but three of streams (streams 1, 3, and 9) were only partly recovered (i.e. not across all regions that the streams span). Streams 1 and 3 are closest to the center of M31 and are located in higher-density backgrounds. The fact that it is the inner streams that are only partially detected is expected due to the lower contrast between the streams and background (see Figure \ref{fig:drho_analytic}, Appendix \ref{sec:analytic}). With a higher Pr-thresh in the \texttt{HSS} run, we could detect these parts of the synthetic streams too, but with the caveat that the code would also find more spurious features. 

For the synthetic streams with 25\% of the 311 initial stream stars, none of the streams are detected in the \texttt{HSS} run, because the synthetic streams are too sparse to stand out against the background. 
This is expected based on the relation between number of stream stars versus log$_{10}$Pr-value (see Appendix \ref{sec:analytic}, Figure \ref{fig:drho_analytic}). We have used a Pr-thresh $<-15$ in the \texttt{HSS}  run, motivated by the fact that several streams are detected on top of one feature if we go lower than this (see Section \ref{sec:bkg}). 
We re-ran the \texttt{HSS} for the 25\% synthetic streams with a Pr-thresh $<-10$ instead.  In this new run, we fully recovered stream 5, located in the low-density outskirts of M31's halo, and we partially rediscovered five of the ten synthetic streams (see Figure \ref{fig:completness_summary}). Four of the streams were not recovered (see ``-'' markers). The \texttt{HSS} also flagged many more unknown (gray) features across the regions (see Appendix \ref{sec:appendix_completeness}). Note that the synthetic streams were flagged as ``blobs'' by the  \texttt{HSS} several times in this run, as multiple detections were found in one region due to the higher Pr-thresh.

From Figure \ref{fig:completness_summary} it is clear that the synthetic streams with a higher percentage of remaining stars have a more significant detection in the \texttt{HSS} (darker colors). We conclude, unsurprisingly, that streams with fewer stars will be detected by the \texttt{HSS} with a larger log$_{10}$Pr-value (less significance). Additionally, if a $10{M}_{\rm Pal 5}$ in M31 only has 25\% of the stars, we will not detect it with an \texttt{HSS}-run with log$_{10}$Pr $<-15$. Thus, in the PAndAS data we are complete for streams with half of the stellar density of a $10 {M}_{\rm Pal 5}$-type stream ($\sim 5{M}_{\rm Pal 5}$-type stream) with the \texttt{HSS}.

As mentioned in Section \ref{sec:grid}, the \texttt{HSS} is capable of detecting thinner, shorter Pal 5--like streams also (i.e. not the $10{M}_{\rm Pal 5}$ used in this section). To check the completeness of Pal 5--like streams in PAndAS, we also inject ten different $2 {M}_{\rm Pal 5}$--like streams \citep[nstars = 68, length = 12 kpc, width = 127 pc; see Table 1 and Figure 1 in][]{pearson19}. We run the \texttt{HSS} with $\Delta\rho = 0.3$ kpc and log$_{10}$Pr $< -10$, and find that the \texttt{HSS} recovers all of the synthetic $2 {M}_{\rm Pal 5}$--like streams. However, the \texttt{HSS} also flags $>$700 features with this threshold, and we get overwhelmed by noise (see also Appendix \ref{sec:appendix_pal5}). When we remove 50\% of the stars and inject ten Pal 5--like streams (nstars = 34, length = 12 kpc, width = 127 pc), only half of the streams are recovered by the run and we again detect more than 700 features. It is possible that some of the $>$700 features are indeed Pal 5--like streams but with the depth of the data, it is not possible to confirm this, and we cannot yet set limits on the presence of Pal 5--like streams in M31. 

\section{Results}\label{sec:results}
We have demonstrated that the \texttt{HSS} can recover synthetic injected streams as well as re-detect the known PAndAS debris features. In this Section, we explore whether the \texttt{HSS} recovers new, unknown GC candidate streams. We first run the  \texttt{HSS} blindly on the PAndAS data after a metallicity cut of [Fe/H] $<-1$ (Section \ref{sec:resrht}), and subsequently analyze the morphology and CMDs of the flagged  \texttt{HSS} candidates (Section \ref{sec:CMD}).

\begin{figure*}
\centerline{\includegraphics[width=\textwidth]{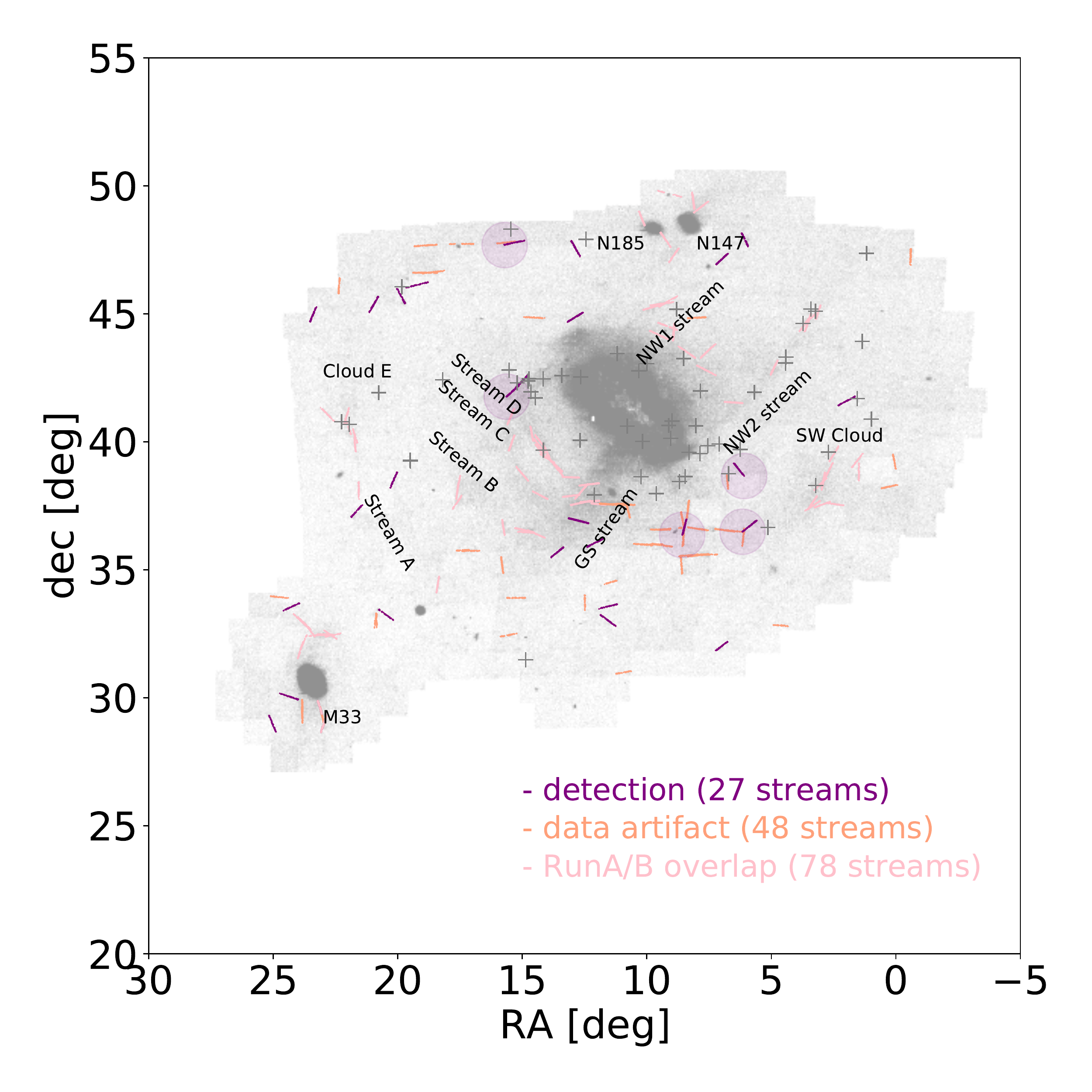}}
\caption{Flagged \texttt{HSS} candidates from a blind run on 2766 overlapping PAndAS data regions (with [Fe/H] $<-1$) and a radius of $r_{\rm angular} = 0.365~ \deg$ with search parameters set to $\Delta \rho = 0.4$ kpc, Pr-thresh $<-15$. The known dwarfs, M31 and known GC were masked out in this run. The \texttt{HSS} flags 153 candidate streams, where 27 streams (purple) are potential new GC candidate streams, and where 48 of these are likely data artifacts (salmon) as $\theta = 0$ or $90$ $\deg$ and trace out the CFHT $1\deg \times1$ $\deg$ fields. The remaining 78 of the flagged streams fall within 0.5 $\deg$ of known dwarf streams found in \texttt{RunA}/\texttt{RunB} (pink). We highlight the five most significant GC candidates (with the lowest log$_{10}$Pr value) with purple transparent circles for further investigation. The gray ``+" markers indicate the location of outer halo GCs in the PAndAS data \citep[][]{huxor14}, where 10 of the GCs are within 0.5 $\deg$ (${\approx}6.9$ kpc) of purple candidate streams. 
Note that several of the purple candidate GC streams are close to dwarf streams and data artifacts, and that none of them trace out several regions as in the idealized case with synthetic data streams in Figure \ref{fig:completness}.}
\label{fig:gc_candidates}
\end{figure*}

\subsection{GC stream candidates in M31}\label{sec:resrht}
We again divide the PAndAS data into 2766 overlapping, equal-area regions with radius $r_{\rm angular} = 0.365~ \deg$ (see Figure \ref{fig:m31}, Sections \ref{sec:data} and \ref{sec:completeness}). Subsequently, we run the \texttt{HSS} on the 2766 regions with a metallicity cut of [Fe/H] $<-1$. 
We have again masked out dwarfs and GCs from PAndAS \citep{huxor14,martin17,McConnachie19} and M31 (see Section \ref{sec:data} and Figure \ref{fig:completness}).
We use parameters optimized for finding $10{M}_{\rm Pal 5}$ synthetic streams: $\Delta \rho = 0.4$ kpc, Pr-thresh $<-15$ (see Section \ref{sec:optimize}). We additionally require that $\theta_{\rm smear} > \frac{\Delta \rho}{2\rho_{\rm max}} > 2.86 \deg$, and $\theta$-separation $>10\deg$ (see Section \ref{sec:bkg}). We remove edge detections where $\rho>3$ kpc, as the 2766 regions overlap by 50\% in both R.A. and decl. (see Section \ref{sec:data}), such that anything on the edge of a region will be in the center of another field. Thus, if edge detections are indeed streams, as opposed to larger overdense features on the edge of a region, they will be flagged as a stream candidate in a separate region. We again exclude detections where ten or more structures were detected in one region, as these are likely ``blobs'', which trace out a full sinusoid in $(\theta,\rho)$-space. While we have masked out dwarfs and GCs from the sample, some ``blobs'' remain  in the data, and these show up as sinusoids in the ($\theta,\rho$) space. While the \texttt{HSS} removes ``blobs'' by flagging a region with more than 10 detections as a ``blob'', there are instances where only part of the sinusoid is above the Pr-thresh. In these instances, part of a sinusoid can be flagged as a stream candidate. If a GC candidate is ``blob"--like in position space or sinusoid-like in ($\theta,\rho$) space, we do not include it in the remainder of our analysis. This was the case for ten flagged detections. 

Of the 2766 regions, the \texttt{HSS} flags stream detections in 153 regions. To investigate which of these 153 flagged detections could be potential GC candidate streams, in Figure \ref{fig:gc_candidates}, we plot the PAndAS data and highlight the stars from most significant \texttt{HSS} detection in each of these 153 regions (i.e. we only plot one detection per region).  We separate the candidates into three groups: (1) streams that are new GC candidates (purple: 27 streams), (2) streams that are likely artifacts of the data as they are at 0 or 90 deg and trace out the CFHT $1\deg \times1\deg$ pointings (salmon: 48 streams), or (3) streams that fall within 0.5 $\deg$ of the detected dwarf features from \texttt{RunA} and \texttt{RunB} (pink: 78 streams; see Section \ref{sec:dwarfs}). 

From Figure \ref{fig:gc_candidates}, we note that none of the detected streams (purple) span several regions, as opposed to the synthetic streams in Figure \ref{fig:completness}, and that several of the purple candidate GC streams appear to be at the edge of an artifact (salmon) indicating that these could be edge detections of these artifacts despite the $\rho_{\rm edge}$-criterion. Several of the dwarf features from \texttt{RunA/RunB} are partially re-detected in this \texttt{HSS} run, which has a narrow $\Delta \rho$-value (see pink streams in Figure \ref{fig:gc_candidates}). This was expected based on the tests in Section \ref{sec:bkg}, where we found that if $\Delta \rho$ was narrower than a specific feature, multiple streams are flagged on top of that feature. We have marked the location of all outer halo GCs \citep[][]{huxor14} with gray ``+" markers in Figure \ref{fig:gc_candidates}. We use the \citet{astropy13, astropy18} SkyCoordinate module to determine that 10 of the GCs are within 0.5 $\deg$ (${\approx}6.9$ kpc) of purple candidate streams, and 17 of the GCs are within 1 $\deg$ (${\approx}13.7$ kpc) of the purple candidate streams.   Several of these candidate streams have orientations that cannot be extrapolated from the GC path (e.g., they are offset perpendicular from the cluster), and are unlikely to be associated with the GCs. Since \citet{huxor14} searched for stream candidates close to the progenitors using HST data, our blind search is more likely to find fully disrupted streams that are not associated with any cluster \citep[see also][]{balbinot18}.

In Appendix \ref{sec:appendix_pal5}, we show the results of an \texttt{HSS} run with $\Delta\rho = 0.3$ kpc and Pr-tresh $< -10$, where we are sensitive to streams with $2 {M}_{\rm Pal 5}$--like streams but also recover $>700$ other features.

\begin{figure*}
\centerline{\includegraphics[width=0.8\textwidth]{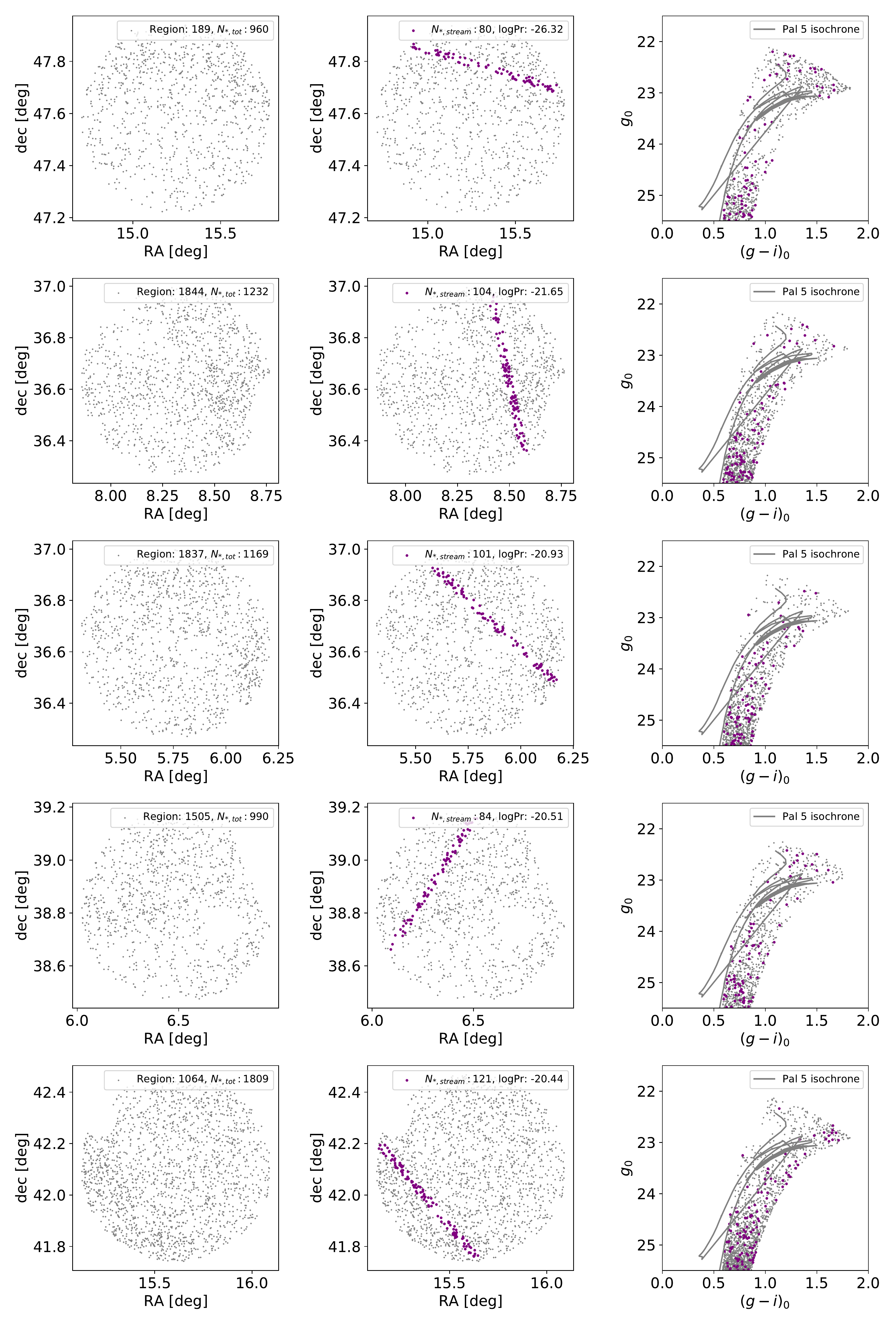}}
\caption{The five most significant GC stream candidate detections from Figure \ref{fig:gc_candidates} (see purple transparent circles).  This figure shows all the stars (gray) in the region  where the detection was flagged in spherical sky coordinates (left column), all the stars as well as the highlighted \texttt{HSS} flagged stream stars (purple) that were within the stripe of detection in spherical sky coordinates (middle). Lastly, we plot the $g_0$ versus $(g-i)_0$ CMD (right column) of all stars in each region (gray), as well as the stars flagged by the \texttt{HSS} as a candidate (purple). We additionally overplot the Pal 5 isochrone (gray line)  with  age = 11.5 Gyr and [Fe/H] $= -1.3$ obtained from the PARSEC (PAdova and TRieste Stellar Evolution Code) set of isochrones \citep{bressan12}, scaled to the distance of M31 (see \citetalias{pearson19} for the shape of the entire isochrone).  Note that we cannot see the main-sequence turn off of a Pal 5--like isochrone at PAndAS' limiting magnitudes at the distance of M31.
}
\label{fig:cmd}
\end{figure*}

\subsection{Morphology and Color-Magnitude Exploration}\label{sec:CMD}
To explore whether any of the 27 purple streams in Figure \ref{fig:gc_candidates} are likely new GC stream candidates, we investigate the morphology of each stream and their location in CMDs. 
We expect that a GC stream would have an old population of metal-poor stars. We therefore compare the CMD of each purple detection to the expectation from an old Pal 5--like globular cluster isochrone (age = 11.5 Gyr, [Fe/H] = $-1.3$) at the distance of M31 (see also Fig. 1 in \citetalias{pearson19}). Note that the main-sequence turnoff is not observable at the limiting magnitude of PAndAS at the distance of M31.

In Figure \ref{fig:cmd}, we highlight the five most significant detections (with minimum log$_{10}$Pr-values) of the purple stream candidates (see purple transparent circles in Figure \ref{fig:gc_candidates}). 
In the left column we plot each star (gray) in the flagged region. In the middle panel, we highlight the stars that were flagged as a \texttt{HSS} detection (purple). In the right column, we plot the $g_0$ versus $(g-i)_0$ CMD for each star in the region (gray), we highlight which stars are in the \texttt{HSS} detection (purple), and overplot the part of the isochrone of a Pal 5--like cluster at the distance of M31 (age = 11.5 Gyr, [Fe/H] = $-1.3$,  gray line). We obtained the isochrone from the PARSEC set of isochrones \citep{bressan12}, which were constructed by interpolating points along missing stellar tracks, which gives rise to the nature of the isochrone's asymptotic giant branch. 
Note that the narrow spread in the $(g-i)_0$ color is due to the nature of how PAndAS photometrically determines metallicities for all stars by assuming that the width of the RGB can be interpreted as the spread in metallicity within a galaxy \citep[see e.g.,][]{denja14}.

\begin{figure}
\centerline{\includegraphics[width=\columnwidth]{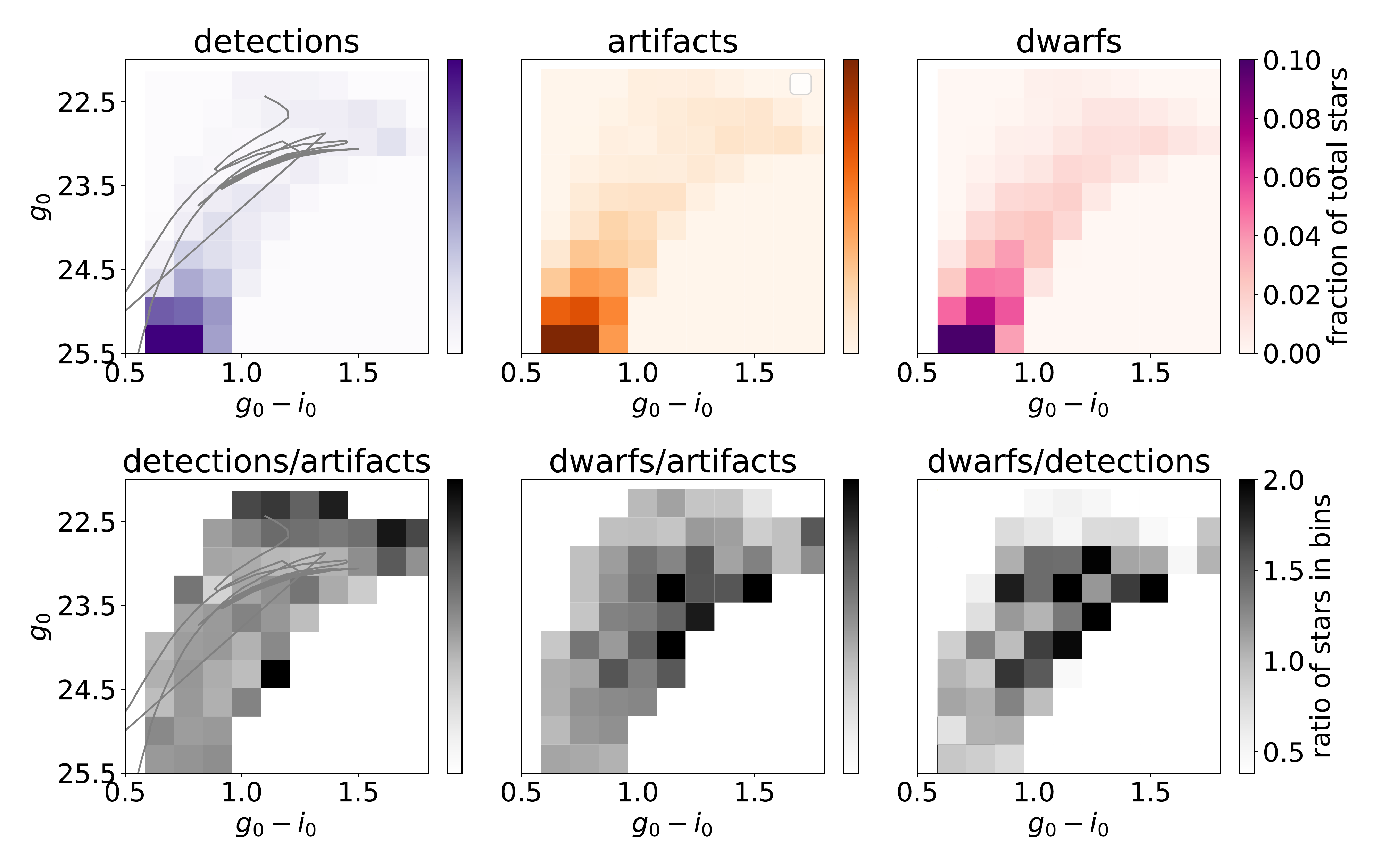}}
\caption{$g_0$ versus $(g-i)_0$ (same as Figure \ref{fig:cmd}, right) but now binned and colored by the fraction of total stars falling within each bin (see color bar) for all of the stars in the 27 candidate GC stream detections (purple: upper left), all of the stars in the 48 flagged artifact (salmon: upper middle), and all of the flagged candidates within 0.5 $\deg$ of known dwarf streams (pink: upper right). We overplot the part of the Pal 5 isochrone (the tip of the RGB) that is visible at the distance of M31 in the upper-left panel (gray). The bottom panel shows the fractional difference between the three types of flagged streams. The GC detections do not show a stronger correlation in the CMDs along the RGB than the artifacts (lower left), although the dwarf streams show more of a signal close to the RGB of a metal-poor isochrone than both the artifacts (lower middle) and the GC candidates (lower right). 
}
\label{fig:cmd_all}
\end{figure}

Some of the stream detections (Figure \ref{fig:cmd}, middle, purple) are noticeable by eye (left). From the right column, it is evident that none of the candidate streams (purple) appear to be strongly clustered around the Pal 5--like isochrone, but instead are scattered in the CMD space. To investigate the GC candidates further, for all of the 27 purple streams in Figure \ref{fig:gc_candidates}, we combine the photometry of all of their constituent stars in the CMD space. In Figure \ref{fig:cmd_all} (upper left), we plot a 2D histogram of stars in $g_0$ versus ($(g-i)_0$ for those 27 streams and color the bins by the fraction of the total number of the stars in those 27 streams that fall within that bin. We carry out the same analysis for all 48 artifact streams (top middle), and all 78 stream candidates that fell within 0.5 $\deg$ of known dwarf features (top right). If the 27 purple streams are indeed real GC candidates from old, low-metallicity globular clusters, we would expect them to have a stronger signal in the CMD along Pal 5's isochrone (see gray line upper left), than the artifacts (top middle), which are not originating from one object. Note, however, that the globular clusters in M31 have a large spread in metallicities \citep[see e.g.,][]{barmby00,caldwell16}, and that streams from younger, more metal-rich globular clusters could show a stronger correlated signal offset to the right of Pal 5's isochrone along the RBG.

\begin{figure}
\centerline{\includegraphics[width=\columnwidth]{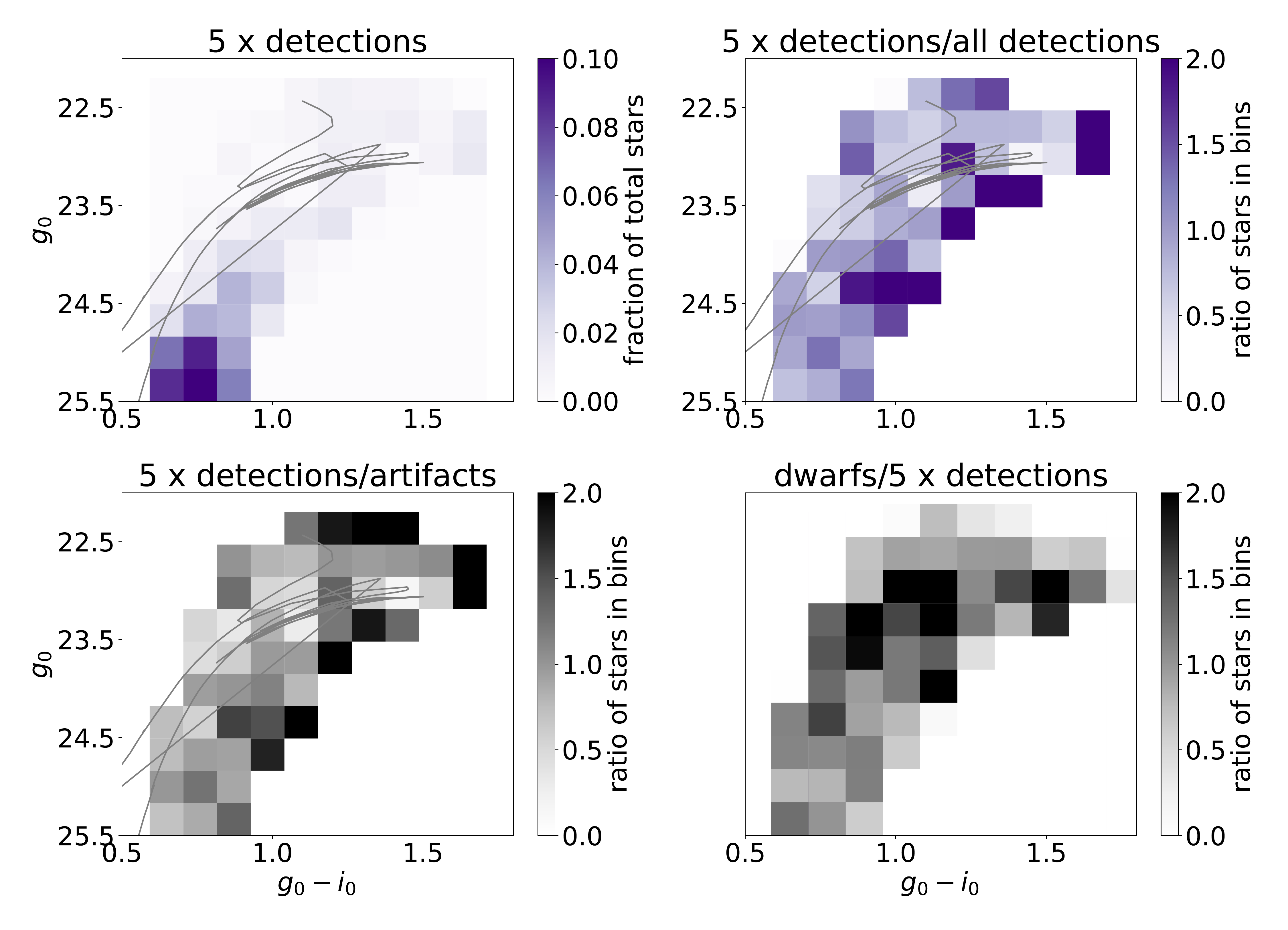}}
\caption{The upper-left panel shows the $g_0$ vs. $(g-i)_0$ binned and colored by the fraction of total stars falling within each bin  (same as Figure \ref{fig:cmd_all}, upper left) but now for the stars in the five most significant GC candidate detections only (see Figure \ref{fig:cmd}). The upper-right panel shows the difference between the 2D histograms for these five most significant candidates vs. the CMDs based on all GC candidate stars (Figure \ref{fig:cmd_all}, upper left). There seems to be more of a signal for the 5 more significant detections in the right part of the CMD (as we saw for the dwarfs in figure \ref{fig:cmd_all}). This does not trace Pal 5's isochrone (gray line). The lower-left  panels show the ratios between the 2D CMD maps for the five most significant detections versus all artifact streams (salmon), and the lower-right panel shows this same ratio for all dwarf streams (pink) versus the five most significant GC candidates. These two maps are the same maps as shown in Figure \ref{fig:cmd_all} lower-left and lower-right, but now using only the five most significant purple detections. Comparing the maps interestingly shows more of a signal along the RGB for the five most significant candidate GC streams, than for all candidate streams.
}
\label{fig:cmd_5sign}
\end{figure}

From the top row of Figure \ref{fig:cmd_all}, we note that the distributions of the fraction of total stars fall in similar regions of the CMD for the GC candidates (top left) and artifacts (top middle) with more of a signal in the bottom-left corner of the plot. For the \texttt{HSS} stream detections that were within 0.5 $\deg$ of dwarf features from \texttt{RunA}/\texttt{RunB} (see pink streams in Figure \ref{fig:gc_candidates}), we notice a slightly stronger signal along the RGB (top-right panel). To investigate this further, in the bottom row of Figure \ref{fig:cmd_all}, we compare the three maps from the top row. In particular, we plot the map of the GC streams (upper left) divided by the map of the artifact streams (upper middle) in the lower left panel to illuminate any differences. Similarly, we plot the dwarf streams map divided by the artifacts map in the lower-middle panel, and the dwarf streams CMD map divided by the GC candidates CMD map in the lower-right panel. 

In the lower-left panel of Figure \ref{fig:cmd_all}, we see that there is no clear difference between the GC candidates and the artifacts (the bin values are $\approx$1), but the dwarf streams have more of a signal along the RGB than both the artifacts (lower middle) and the GC candidate (lower right) where the values of the map ratios are $>1$. As the streams on top of the dwarf features (pink) are expected to trace metal-poor populations of disrupted dwarfs, it is not strange that these show correlated colors in the CMD. However, if the purple \texttt{HSS} detections are indeed true GC streams, we expect them to show this same trend of a stronger signal along the RGB in the CMD than for the artifacts. This could be along slightly different isochrones than the Pal 5 isochrone overplotted here (e.g., a more metal-rich GC would have an isochrone to the right of the Pal 5--like isochrone in the $g_0-i_0$ color).

We now redo the analysis above, but use only the stars from the five most significant detections  (see purple circles on Figure \ref{fig:gc_candidates} and the highlighted streams in Figure \ref{fig:cmd}) to create our binned maps showing the 2D histogram of the location of all stars in those five streams. We again color the bins by the fraction of the total number of the stars in those five stream candidates that fall in each bin. We show this map in Figure \ref{fig:cmd_5sign} (upper left). We compare this new map to the previous map of all 27 candidate purple streams (Figure \ref{fig:cmd}, upper left) by dividing the 2D histogram of the five most significant detections with the  2D-histogram of all 27 detections (see Figure \ref{fig:cmd_5sign}, upper right). There is a stronger signal along the RGB for the five most significant detections than in all 27 candidates as the values are $>1$ (see color bar); however, this enhancement does not fall exactly along the Pal 5--like isochrone. This could indicate that these candidates are streams from younger, more metal-rich globular clusters \citep[see GC metallicity distributions for M31 in, e.g.,][]{barmby00,caldwell16}.

In the lower-left panel of Figure \ref{fig:cmd_5sign}, we compare the new fractional 2D map of the five candidate streams (Figure \ref{fig:cmd_5sign}, upper left) to the fractional 2D map of all 48 artifact streams, which is the same map that we showed for all purple streams in the lower-left panel of Figure \ref{fig:cmd_all}. In the lower-right panel of Figure \ref{fig:cmd_5sign}, we show the fractional 2D map of all stars in the 78 streams that fell in the vicinity of dwarf streams divided by the map of all stars in the five GC candidate streams (Figure  \ref{fig:cmd_5sign}, upper left). Interestingly, the maps in Figure \ref{fig:cmd_5sign}, using the five most significant streams, differ from the maps presented in Figure \ref{fig:cmd_all}, where we used all purple GC stream detections from Figure \ref{fig:gc_candidates}. In particular, there appears to be a stronger signal (values $>1$) along the RGB for the five CG candidates compared to the artifacts (Figure \ref{fig:cmd_5sign}, lower left), and there is a shift in the distribution of the dwarfs versus candidates (Figure \ref{fig:cmd_5sign}, lower right). However, there is still a stronger signal in the dwarf streams along the Pal 5--like isochrone than for the five GC stream candidates.

The fact that the 27 GC candidates (Fig \ref{fig:gc_candidates}, purple) do not appear to be correlated in the CMDs nor follow a Pal 5--like isochrone could indicate that these flagged detections are noise or artifacts as well. However, we do see a higher signal along the RGB in the CMD when we analyze the most significant candidates only (see Figure \ref{fig:cmd_5sign}). 
The data appear to be just at the boundary, where it is difficult to validate (and detect) GC streams in the PAndAS data. We discuss this further in Sections \ref{sec:newdiscussion} and \ref{sec:discussion}.

\section{Discussion}\label{sec:newdiscussion}
In this Section, we discuss our choice of parameters for the blind \texttt{HSS} run (Section \ref{sec:performance}), and how the \texttt{HSS} code compares to and differs from other existing stream-finding techniques (Section \ref{sec:othercodes}).

\subsection{The HSS parameter choices}\label{sec:performance}
In Section \ref{sec:results}, we showed the results of a blind \texttt{HSS} run with a specific choice of parameters, such as the significance of detection threshold (Pr-thresh $<-15$), search with ($\Delta\rho = 0.4$ kpc), region size ($d_{\rm ang} = 0.73~ \deg$), and metallicity cut ([Fe/H] $< -1$). The choice of Pr-thresh $<-15$ and $\Delta\rho = 0.4$ kpc were motivated in Section \ref{sec:optimize} based on a $10{M}_{\rm Pal 5}$ synthetic stream. Specifically, Pr-thresh $<-15$ lead to the detection of a $10{M}_{\rm Pal 5}$ synthetic stream without adding noisy spurious detections (see also Appendix \ref{sec:appendix_completeness}).  $\Delta\rho = 0.4$ kpc optimized the detection of the synthetic stream, while streams narrower than the chosen $\Delta\rho$ value could still be flagged as stream overdensities, but with a lower significance. 

If we change the region size and the exact region location, this can change the specific detections that are closer to the edges of our regions, which still pass the $\rho_{\rm edge}$-cut (see Section \ref{sec:bkg}). Throughout this work, we have ensured that the region sizes are at least ten times larger than the feature we search for, such that stream--like features would not fill up an entire region and go undetected. We have additionally used overlapping regions, such that a feature at the edge of one region will be at the center of its neighboring region. 
While a different choice of region locations might change the specific flagged data artifacts, we do not expect true stream candidates to be affected by a shift in region locations and sizes (see example of how the random injection of synthetic streams lead to clear \texttt{HSS} detections in Section \ref{sec:completeness}). 

Lastly, we can use a less restrictive metallicity cut, as some GCs in M31 have higher metallicities and younger ages than Pal 5 \citep[e.g.,][]{caldwell16}. To test the effect of a less restrictive metallicity cut, we re-run the \texttt{HSS} with the same parameters as in Section \ref{sec:results}, but on PAndAS data with [Fe/H] $< 0$. Dynamically, it takes GC streams several gigayears to form and evolve \citep[e.g.,][]{johnston01}, and thus adding in more stars with higher metallicities could serve to contaminate our sample further with non-stream members. In this new run, the \texttt{HSS} flagged more candidate streams (70), data artifacts (72), and more streams within 0.5 $\deg$ of the dwarf debris (90) as compared to the \texttt{HSS} run with [Fe/H] $< -1$ presented in Figure \ref{fig:gc_candidates}. This is not surprising, as adding more stars to the regions while keeping Pr-thresh fixed should lead to higher significance detections (see Appendix \ref{sec:analytic}). Of the 27 GC stream candidates found in the [Fe/H] $= -1$ run (see Section \ref{sec:results}), 20 of those candidates are within $0.5\deg$ of the 70 new candidate streams from the [Fe/H] $= 0$ run.
With the less conservative metallicity cut, we again found that there is more of a signal in the dwarf streams in the RGB part of the CMD than for the artifacts and GC candidates, and it is difficult to make conclusive statements about the candidates. 

Thus, we can detect more GC candidates by re-running the \texttt{HSS} with a grid of parameters. However, as the PAndAS data appear to be at the very limit of detection capability for GC streams in M31, we leave this exploration for future analysis of deeper data.

\subsection{Other stream-finding techniques}\label{sec:othercodes}
We have presented a new technique to identify streams in resolved stars, and we can quantify how much of an outlier our stream detection is with respect to its background. 
Over the past few decades, different techniques have been developed to search for stellar streams in various multidimensional data sets, but to date, there is still no universal way of confirming the significance of a stream detection. 
In this section, we discuss the current state-of-the-art for stream-finding, most of which relies on follow-up measurements of kinematics, colors, and metallicities in addition to positional information.

One of the first examples of stream-detection methods was 
presented in \citet{johnston96}, who showed that debris structure from tidally disrupted satellites could remain aligned along great circles passing close to the Galactic poles for several gigayears. They developed a method named the Great Circle Cell Counts (GC3), where they create a grid of great circle cells with equally spaced poles to provide a systematic search for debris trails along all possible great circles. GC3 was used to identify the first full sky view of the stream from the Sagittarius dwarf \citep{ibata02b,majewski03}. As streams are not only coherent linear structures in positional space, but also exhibit ordered kinematics, \citet{mateu17} has since extended the GC3 method to include kinematic information. 

Individual streams likely originate from just one progenitor, which means that streams, most often, consist of a specific population of stars with distinct ages and metallicities. Dwarf streams can consist of several populations and will therefore have a larger spread in these quantities. In the early days of stream finding, \citet{grillmair95} took advantage of the fact that the progenitor main sequence on the CMD should be the same for extended tidal debris near globular clusters. Since then, this technique has been expanded and optimized to find specific stellar populations against noisy stellar backgrounds through matched-filter techniques in color-magnitude space \citep[e.g.,][]{Rockosi02}. The matched filtering technique relies on a `template isochrone' of a specific age and metallicity representing a specific population of interest. Similarly, the technique relies on knowledge of a ``background template", such that it is possible to construct a weighting filter that maximizes the signal-to-noise of the output map. With these templates in hand, it is possible to select a range of stars around the template isochrone within the CMD while stepping in distance modulus. Several groups have found numerous MW streams using this technique \citep[e.g.,][]{grillmair06,bonaca12,carlberg12,ibata16,Shipp18,ship20,thomas20}. 

With the wealth of stellar halo data that will be available in the near future from Roman \citep[][]{spergel15}, VRO \citep[][]{laureijs11}, and Euclid \citep[][]{racca16}, it will be important that we do not only detect substructure, but also classify detections of various substructure such as streams and shells to learn astrophysical parameters from their properties.
\citet{hendel19} developed a machine vision method, SCUDS, which automates the classification of debris structures (see \citealt{Darragh-Ford2020} for a dwarf-finding algorithm). In particular, the algorithm first locates high-density ``ridges'' that are typical of substructure morphology in  controlled N–body simulations of minor mergers. Once a `ridge' has been located, the algorithm determines whether it is ``stream''--like or ``shell''--like based on an analysis of the coefficients of an orthogonal series density estimator. With SCUDS applied to current \citep[e.g.,][]{delgado10,delgado19,delgado21} and future large data sets, we will be able to obtain global morphological classifications, which will help statistically assess, e.g., the host-to-satellite mass ratio, the interaction time, and the satellite orbits for a large sample of galaxies.

\citet{malhan18a} developed the code  \texttt{STREAM-\\FINDER} where their aim was to use maximal prior knowledge of stellar streams, including kinematics, to maximize the detection efficiency. Their code makes use of 6D hyperdimensional (position and velocity) ``stripes'' in phase-space, with plausible widths and lengths motivated from a disrupted progenitor's properties and orbit. By integrating trial orbits and searching within 6D hyperdimensional ``tubes'' surrounding these orbits, \texttt{STREAMFINDER} identified several known streams and new stellar stream candidates \citep[][]{malhan18b,ibata19b} in Gaia DR2 data \citep[][]{gaiadr2,Lindegren18}, most of which have since been confirmed via the coherence in  their radial velocities \citep[][]{ibata21}.
Recently, \citet{shih21} applied a data-driven, unsupervised machine-learning algorithm, ANODE \citep{nachman20}, which uses conditional probability density estimation to identify anomalous data points along with a Hough transform, to search for streams in Gaia DR2 data \citep[]{gaiadr2}. In particular, they identify the region in Hough space with the highest contrast in density compared to the region surrounding it, and search the Hough space for the parameters that maximize the significance of their detection. The input for the ANODE training includes the angular position, proper motion, and photometry of the stars, which is ideal for data sets such as Gaia DR2 \citep[]{gaiadr2}. 
However, in external galaxies, we will most often not have access to kinematic data, and ``blind'' systematical, morphological searches (such as carried out by \texttt{HSS}) will be critical.

The \texttt{HSS} is developed with external galaxies and future surveys of resolved stars in mind, and it currently uses positional information only. In contrast, the GC3 method \citep[][]{johnston96} was built for an internal Galactic perspective.
We expect \texttt{HSS} to be a great tool to rapidly and systematically identify streams in densely populated data sets of resolved stars. Due to \texttt{HSS}'s general 
nature, its application is not limited to searches for stellar streams, but could be adapted to search for linear structure in other data sets. Similarly, the  \texttt{HSS} can be extended to include color information instead of having this as a post-processing step.

\section{Future prospects}\label{sec:discussion}
With the \texttt{HSS}, we have found 27 GC stream candidates in PAndAS, but we could not make conclusive statements regarding their nature. In this Section, we discuss the expected GC stream population in M31 and future data that can be used to search for and/or confirm GC stream candidates in M31 (Section \ref{sec:m31vsMW}). We also discuss how the \texttt{HSS} combined with Roman will help find GC streams in external galaxies (Section \ref{sec:rst}), and how this can potentially help in the quest for the nature of dark matter (Section \ref{sec:dm}). 

\subsection{GC population of M31 versus MW}\label{sec:m31vsMW}
The accretion histories of the MW and M31 have differed substantially \citep[e.g.,][]{deason13,mackey19,mackey19b}. We see evidence of this, in part, from the large dissimilarity in the number of GCs orbiting each of the spiral galaxies. While we know of $\approx$150 GCs in the MW \citep[e.g.,][]{harris96}, there are more than 450 reported detections of GCs in M31 \citep[][]{huxor14,caldwell16,mackey19}. In the MW, $<20\%$ of the known GCs show hints of tidal debris surrounding them \citep[e.g.,][]{leon2000,kundu19} with only a few clear examples of extended stellar streams \citep[e.g.,][]{oden01,grillmair06b,ship20}. However, several stellar streams in the MW have been detected in the absence of a progenitor. The initial progenitors of those streams have likely been fully torn apart by tides from the MW's gravitational field. 
Based on the widths and metallicities of the MW streams, $>$50 of them likely originated from disrupted GCs \citep[e.g.,][]{mateu18}. Since M31 has three times more GCs than the MW, it is reasonable to expect that M31 hosts $>150$ GC streams (three tim$10{M}_{\rm Pal 5}$ es the amount of GC streams than the MW), most of which should have fully disrupted progenitors. 

Our work in this paper and in \citetalias{pearson19} has demonstrated that we should be able to detect $5{M}_{\rm Pal 5}$ and $10{M}_{\rm Pal 5}$ streams in the PAndAS data if those streams exist in M31, but that a Pal 5--like stream cannot be detected (see Section \ref{sec:completeness}). However, we did not find clear evidence (e.g., as compared to the synthetic streams in Section \ref{sec:completeness}) of GC stream in the PAndAS data with a systematic search using the \texttt{HSS}. Thus, it appears as though there are no $10{M}_{\rm Pal 5}$ streams orbiting M31, as these should have been detected with a log$_{10}$Pr-value  ${\approx}-80$ (see Figure \ref{fig:completness_summary}), where our most significant detections have a log$_{10}$Pr-value ${\approx}-25$ (see Figure \ref{fig:cmd}, middle columns). It is possible that streams with 50\% of the surface density of a $10{\rm M}_{\rm Pal 5}$--like stream exist in the data (see log$_{10}$Pr-values in Figure  \ref{fig:completness_summary} versus Figure  \ref{fig:cmd}, middle panels) or streams with younger, more metal-rich stars than Pal 5 (see Figure \ref{fig:cmd_5sign}).

The MW stellar stream, Pal 5, had an initial mass of $\approx$47,000 $\pm~ 1500~ \msun$ \citep{ibata17}, and many GCs exist that are much more massive than Pal 5 \citep[see e.g.,][]{harris96,ibata19}. Thus, it is not unreasonable to expect that GC streams that are five to ten times more massive than Pal 5 can exist in M31. Pal 5 has $\approx8,000~ \msun$ in its tails at present day \citep{ibata16,ibata17}. The globular cluster MW stream, GD-1, has ${\approx}2 \times 10^{4}~ \msun$ in its stream \citep[][]{koposov10}, which makes it ${\approx}2.5~ \times$ more massive than Pal 5. In the MW, we do not yet know of GC streams more massive than GD-1 at present day, and it might be that there simply are not any GC streams that massive in M31's stellar halo. Note, however, that many factors play into our ability to detect such streams in the MW (e.g., location in the Galaxy, time of accretion, and extinction).

If the M31 stream mass population is similar to the observed MW stream population, the PAndAS data appear to be at the very boundary of detection capability for GC streams. 
To get a better probe of the CMDs for the GC candidate structures and to confirm the nature of the GC candidates in this work, deeper data is needed. This would allow us to probe the RGB of potential old GC streams in M31's stellar halo. Deeper surveys could also find a wealth of one to two times Pal 5--like streams. The HST, and soon the James Webb Space Telescope, and its Near Infrared Camera, are ideal for deeper data \citep[e.g.,][]{huxor14}, but due to the small FOV, they are not ideal for anything spanning more than a few arcminutes (i.e. much smaller than the synthetic streams in this work which span $1.89 \deg$). Thus, this would be an expensive and risky observational program.  

Interestingly, \citet{patel18} showed that the Hyper Suprime-Cam (HSC) on Subaru, which has a much larger FOV of 1.8 deg$^2$, can go 1.5 mag deeper than PAndAS. However, it is unclear how well the HSC will be able to resolve enough of the individual stars from the much more numerous unresolved background galaxies at faint magnitudes. Similar surveys can be done with Magellan+Megacam, but with much smaller FOVs. Instead, future surveys carried out with wide field telescopes that resolve individual stars, such as Roman, are perfectly suited for this purpose (\citetalias{pearson19}), and when they are combined with the \texttt{HSS} we can carry out a systematic search. 

\subsection{\texttt{HSS} and Roman Space Telescope Synergy}\label{sec:rst}
With the large FOV (0.28 $\deg{^2}$) and high spatial resolution (0.11$\arcsec$) of the Nancy Grace Roman Space Telescope (Roman), we know that GC streams can easily be resolved and stand out against the background of M31  (\citetalias{pearson19}). The HST PHAT (the Panchromatic Hubble Andromeda Treasury) survey \citep{phat12} used 432 pointings to cover the disk of M31. This can be done using only two pointings with Roman. For comparison, the entire field of PAndAS (400 deg${^2}$), can be covered in $\approx$1500 pointings, but have a similar spatial resolution and depth as the HST. While such a program is not yet planned, the fact that a large part of GC streams can be covered in one Roman pointing makes Roman ideal for follow-up to verify and characterize, e.g., HSC candidates. 
In this Section, we demonstrate the ability of \texttt{HSS} to find Pal 5 in future 1 hr exposure Roman data of M31's stellar halo.

We inject a stream with the present-day mass of Pal 5's stream \citep[i.e., not with ten times the mass]{ibata17} to a background M31 field, which represents Roman's limiting magnitudes and stellar densities at a galactocentric distance of $R_{gc} = 55$ kpc (see sec. 3.1.2 in \citetalias{pearson19}). The length of the stream is updated based on the gravitational potential of M31 at a galactocentric radius of 55 kpc. At this location in M31, Pal 5 would have a width of $w = 0.127$ kpc, and would have 1299 resolved stars based on the limiting magnitude of a 1 hr Roman exposure at the distance of M31 (see Figure 1 and Table 1 in \citetalias{pearson19}). 

We inject the stream to a region size representing Roman's FOV (i.e. $r_{\rm angular} = \sqrt{0.28\deg} = 0.529 \deg$. We therefore use a radius of $0.529/2 \deg$ in this example, which is $\approx 3.62$ kpc at the distance of M31).  In this example, we apply a metallicity cut of [Fe/H]$<-1$, and we run \texttt{HSS} with $\Delta \rho = 0.3$ kpc, and  $\theta_{\rm smear} = \frac{\Delta \rho}{2 \times \rho_{\rm max}} > \frac{0.3~ {\rm {\rm kpc}}}{2\times3.62~ {\rm kpc}}= 0.041~ {\rm rad} = 2.37 \deg$ (see details in Sections \ref{sec:grid} and \ref{sec:bkg}). 

\begin{figure*}
\centerline{\includegraphics[width=\textwidth]{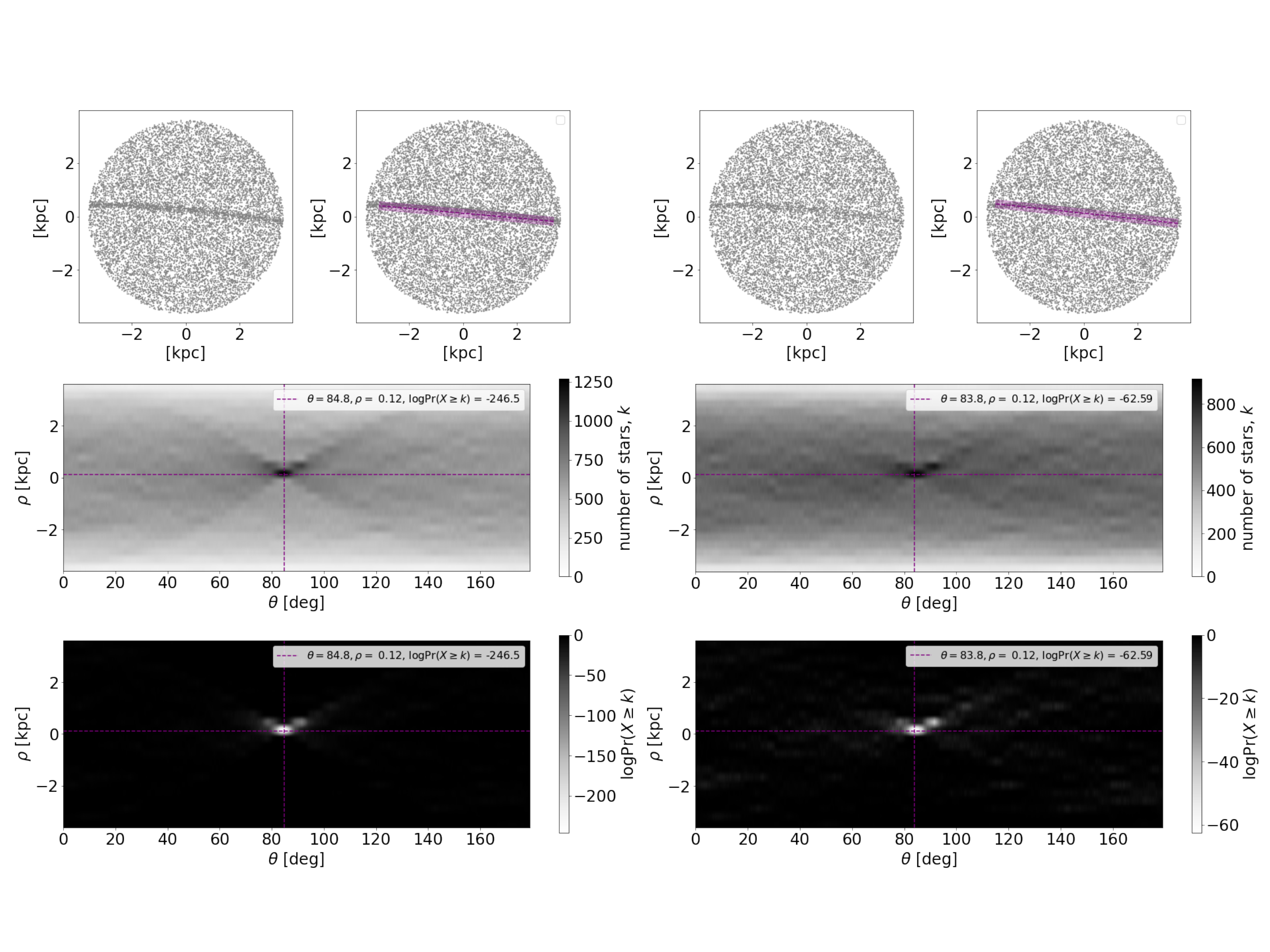}}
\caption{Top panels: input data to \texttt{HSS} with resolved stars in M31 background with Roman limiting magnitudes and a Pal 5 stream (a) and a stream with 50\% of Pal 5 (b) injected with an updated number of stars for a limiting magnitude of 1 hr Roman exposure at the distance of M31. The purple stripes show the streams detected by the \texttt{HSS}.  Middle panels:  the Hough transform of each star from the top panel shown in a $(\theta,\rho)$-grid where each bin has a certain number of stars, $k$, corresponding to how many sinusoidal curves crossed this bin, which was $\approx 1250$ stars for the Pal 5--like stream (a) and ${\approx}800$ stars for the stream with 50\% of Pal 5 stars (b).
Lower panels:  the probability of the $(\theta,\rho)$-grid (middle) having $k$ or more stars cross each specific bin, by chance (Eq. \ref{eq:binomial}). The purple dashed lines highlight the flagged stream detection (corresponding to the purple ‘stripe’ in the top panels). Note that the probability distribution for case b (lower right) is slightly more noisy. Roman combined with \texttt{HSS} will thus allow us to detect GC streams much fainter than Pal 5 in hundreds of galaxies.
}
\label{fig:wfirst}
\end{figure*}

In Figure \ref{fig:wfirst}, we show the results of \texttt{HSS} run on Roman--like data with: (a) an injected Pal 5--like stream, and (b) an injected Pal 5--like stream with 50\% of the surface density. Note that the length of the stream is much larger ($l=12$ kpc, see Table 1 in \citetalias{pearson19}) than the size of the region (based on Roman's FOV), so the streams will connect over several regions that would be detected by the \texttt{HSS} (see e.g., Fig. \ref{fig:completness}). Note also that due to the stream's larger length, not all 1299 stars are included in this region.  
In the upper panels of of Figure \ref{fig:wfirst}, we show the input data fed to \texttt{HSS}, as well as the recovered stream (purple stripe). The middle panels show the $(\theta,\rho)$ grid, which is the Hough transform of each star (Eq. \ref{eq:HT}) binned in $\Delta \rho$. The gray scale demonstrates how many stars, $k$, fell in each specific bin. The lower panel shows the binomial log$_{10}$Pr($X \geq k$), where $k$ is the value (number of stars) in each bin in the ($\theta,\rho$) grid (middle). The probability of one star landing in a certain bin is represented by $p = dA/A$ (see Eq. \ref{eq:dA}). 

\begin{deluxetable}{lccc}
\tablecaption{Summary of \texttt{HSS} Roman Pal 5--like stream recovery.}
\tablecolumns{3}
\tablenum{2}\label{tab:roman}
\tablewidth{0pt}
\tablehead{\colhead{Remaining Stars} &
\colhead{$\rho$} &
\colhead{$\theta$} &
\colhead{log$_{10}$Pr} \\
\colhead{} &
\colhead{(kpc)} &
\colhead{($\deg$)} &
\colhead{} 
}
\startdata
100\%                    & 0.12 &  84.8 & $-246.5$\\
50\%                &0.12 &   83.8 & $-62.59$ \\
25\%                   & 0.12 & 84.8 & $-23.12$\\
10\%                 & - &  - &$-$ \\
\enddata

\end{deluxetable}

The \texttt{HSS} clearly detects the synthetic Pal 5--like stream (Figure \ref{fig:wfirst}, left).
When we remove 50\% of the stars in the Pal 5--like stream (Figure \ref{fig:wfirst}, right), we still detect the stream, but at a slightly different angle and with a lower significance (see Figure \ref{fig:wfirst}, bottom panels, and a summary in Table \ref{tab:roman}).
When we only include 25\% of the stars in the Pal 5--like stream, the \texttt{HSS} still detects the streams but with orders-of-magnitude lower significance and at a slightly different angle (see Table \ref{tab:roman}), as there is more noise in the surrounding region. We do not detect the stream with 10\% of the stars remaining. 
The fact that our code has the ability to significantly detect streams with 50\% of Pal 5's stars in M31 yields very promising prospects for future GC stream searches with Roman and \texttt{HSS}, in M31 and beyond. While using conservative limits for star-galaxy separation, \citetalias{pearson19} showed that with a 1 hr Roman exposure, we can easily detect Pal 5--like streams within 1.1 Mpc by eye (see their Figure 4, left panel). We will additionally probe three magnitudes down the RGB as compared to the PAndAS data presented here. Thus with Roman and the \texttt{HSS}, we can place PAndAS-quality constraints on GC streams for any halo within 10 Mpc, and Pal 5--like streams can be detected in a wealth of galaxies in the near future.

\subsection{Constraining the nature of dark matter}\label{sec:dm}
Over the next  decade,  there  are  multiple prospects for deploying the \texttt{HSS} to data sets from other galaxies in search of thin GC streams. 
With Roman and \texttt{HSS}, we will have the ability to detect hundreds of these types of streams in the Local Volume of galaxies \citep{kara19,pearson19}, and many in M31 alone. 
This will usher in a new era of statistical analysis of stellar stream morphology. The mass spectrum of dark matter subhalos varies depending on the nature of the dark matter particle \citep[e.g.,][]{bohm14}. If dark matter is indeed a weakly interactive massive particle, as is the case in $\Lambda$ cold dark matter, subhalos with masses lower than 10$^{6}$ $\msun$ should be abundant in galaxies. If dark matter is instead composed of warm, lighter particles, these low-mass subhalos should not exist \citep[e.g.,][]{bullock17}. Thus, finding evidence of the existence of low-mass dark matter subhalos is of particular interest, as that will enables us to distinguish between dark matter particle candidates. 

The thin MW stellar stream GD-1, which is likely the remnant of a fully disrupted GC \citep{grillmair06}, is a prime example of a stellar stream that has a noticeable gap. GD-1 orbits the MW retrograde with respect to the disk and Galactic bar, which means that the stream will be minimally impacted by these components of the galaxy \citep[e.g.,][]{hattori16}. Thus, the gap in GD-1 could be evidence of an interaction between a low-mass dark matter subhalo and a GC stream \citep{boer18,price18,bonaca19,bonaca20b}.
Another example of a MW stream with gaps is the stream associated with Pal 5, which also shows evidence of direction variations in its stream track  \citep[e.g.,][]{bonaca20}. However, due to Pal 5's prograde orbit with respect to the Galactic bar \citep[][]{pearson17, erkal17}, it is more difficult to disentangle the origin of the gaps and morphological disturbances in Pal 5. In external galaxies, we will most often not have access to kinematic data of the stream members, which could make it difficult to put together conclusive claims about the nature of the gaps \citep[see e.g., the complex parameter space of GD-1's perturber in the MW in][]{bonaca19,bonaca20b}. 
On the other hand, if we have a large enough sample of gaps in streams in external galaxies without molecular clouds, bars, and spiral arms, there are less opportunities for baryonic perturbers to induce gaps in streams \citep[][]{amorisco16,hattori16,pearson17,erkal17,banik19,pearson19,bonaca20}. We can potentially also do statistical analyses on gap distributions in streams as a function of environment and galactocentric radii, as smaller subhalos should be destroyed in the central parts of galaxies \citep[][]{kimmel17}. 

The \texttt{HSS} can systematically search for GC streams in future Roman data sets. Although the code is not yet optimized to find gaps, with resolved stars, discontinuous structures should be quite easy to see in post-processing, and we plan to facilitate the search for discontinuities in streams (i.e. gaps) in a future release of the \texttt{HSS} code.

\section{Conclusion}\label{sec:conclusion}
We have developed a new code, the \texttt{Hough Stream Spotter} (\texttt{HSS}), optimized to find and characterize linear structure in discrete data sets. The \texttt{HSS} takes two positional coordinates as inputs, and searches for overdensities via a Hough transform and a binomial probability analysis to flag potential stream candidates in noisy background regions. We have optimized the code to be sensitive to thin GC--like streams through both numerical and analytic analyses of various synthetic streams injected with different number densities and widths, at various locations and orientations. Additionally, we have tested and applied our code to the photometric PAndAS data from M31's stellar halo, and we found the following:

\begin{itemize}

    \item The \texttt{HSS} rediscovers all previously known dwarf streams and clouds in M31's stellar halo, except for the GS, which is flagged as a ``blob" due to our search criteria. The \texttt{HSS} also detects linear artifacts in the data and edges of real features. 
    
    \item The \texttt{HSS} easily detects $10{M}_{\rm Pal 5}$ synthetic streams injected to the PAndAS data. The code traces the synthetic streams' curvatures, and is complete to streams with 50\% of the surface density of a $10 {\rm M}_{\rm Pal 5}$--like stream in M31's stellar halo.
    
    \item We found 27 new GC candidate streams that passed our detection criteria motivated from the synthetic GC streams. The five most significant detections show a stronger signal along the RGB than artifacts in the data, but we need follow-up data to confirm whether they are true GC streams. 
    
    \item We have demonstrated that the Roman Space Telescope will be sensitive to GC streams, and that the \texttt{HSS} can find these streams. Roman and morphology-based codes like the \texttt{HSS} will usher in a new era of statistical analyses of extragalactic GC stream morphologies.

\end{itemize}

While we do not yet have a confirmation of a GC stream in any other galaxy than the MW, there are exciting prospects for using the morphology of GC streams in external galaxies for orbit mapping, potential mapping, and statistical gap assessment in the near future with the Roman Space Telescope. 

\begin{acknowledgements}
We thank Adrian Price-Whelan, David W. Hogg, the Astronomical Data Group, and the Galactic Dynamics Group at the Flatiron Institute for insightful discussions. Support for this work was provided by NASA through the NASA Hubble Fellowship grant \#HST-HF2-51466.001-A awarded by the Space Telescope Science Institute, which is operated by the Association of Universities for Research in Astronomy, Incorporated, under NASA contract NAS5-26555. The Flatiron Institute is supported by the Simons Foundation. S.E.C. acknowledges support by the Friends of the Institute for Advanced Study Membership. K.V.J.'s contributions were inspired by the WFIRST Infrared Nearby Galaxies Survey collaboration and supported in part by NASA grant NNG16PJ28C through subcontract from the University of Washington, as wells a grant form the National Science Foundation, AST-1715582. R.I. acknowledges support from the European Research Council (ERC) under the European Unions Horizon 2020 research and innovation programme (grant agreement No. 834148).
\end{acknowledgements}

\software{
    \package{Astropy}~ \citep{astropy13,astropy18}, 
    ~\package{matplotlib}~ \citep{hunter07}, 
    ~\package{numpy}~ \citep{numpy:2020}, 
    ~\package{scipy}~ \citep{Virtanen:2020}.
}



\appendix
\section{Analytic investigation of HSS parameters for streams}\label{sec:analytic}
In Figure \ref{fig:drho}, the optimal $\Delta \rho$ search width ($\Delta \rho = 0.4$ kpc)  was a bit larger than the stream width in this example ($w = 0.273$ kpc). However, the specific background and number of number of stars in a stream can also affect the detection of a stream. To generalize which search width and Pr-thresh value to use for a specific type of stream search, we investigate a range of stream widths and backgrounds analytically.

We can calculate the binomial probability (see Equation \ref{eq:binomial}) of a detection having ``$k$" or more stars analytically (see Section \ref{sec:bkg}) for an idealized case, where the stream passes through the center of the given region. We can calculate this probability as a function of the stream width, $w$, the number of stars in the stream, the stripe width, $\Delta \rho$, and the number of stars in the background and the area of background. We use an array of $\Delta \rho$ from 0.05--1.5 kpc in steps of 0.05 kpc. In the binomial calculation, $p$ is the same as described in Section \ref{sec:bkg} and is the area of the stripe (which depends on the chosen $\Delta \rho$) divided by the area of the total region: $p = \frac{A_{\rm stripe}}{A_{\rm background}}$. $N_{\rm stars}$ is the total number of stars in the region, $N_{\rm stars}$ = $N_{\rm stream}$ + $N_{\rm background}$, and the number of stars that fall into a certain stripe, $k$, depends on whether the stripe width ($\Delta \rho$) is greater than (gt), or less than (lt) the stream width. If the stripe width is greater than the stream width, all stream stars and the background stars that fall within the stripe will be counted:

\begin{equation}
k_{1_{gt}} = N_{\rm stream} + p N_{\rm background}.
\label{eq:klt_analytic}
\end{equation}
But if the stripe width is less than the stream width, only part of the stream stars will be counted as well as the  background stars that fall within the stripe:
\begin{equation}
k_{2_{gt}} = N_{\rm stream} \frac{A_{\rm stripe}}{A_{\rm stream}} + p N_{\rm background}.
\label{eq:kgt_analytic}
\end{equation}
With this prescription, we can analytically calculate the probability of a stripe having $k$ or more stars via Equation \ref{eq:binomial}.

\begin{figure*}
\centerline{\includegraphics[width=\textwidth]{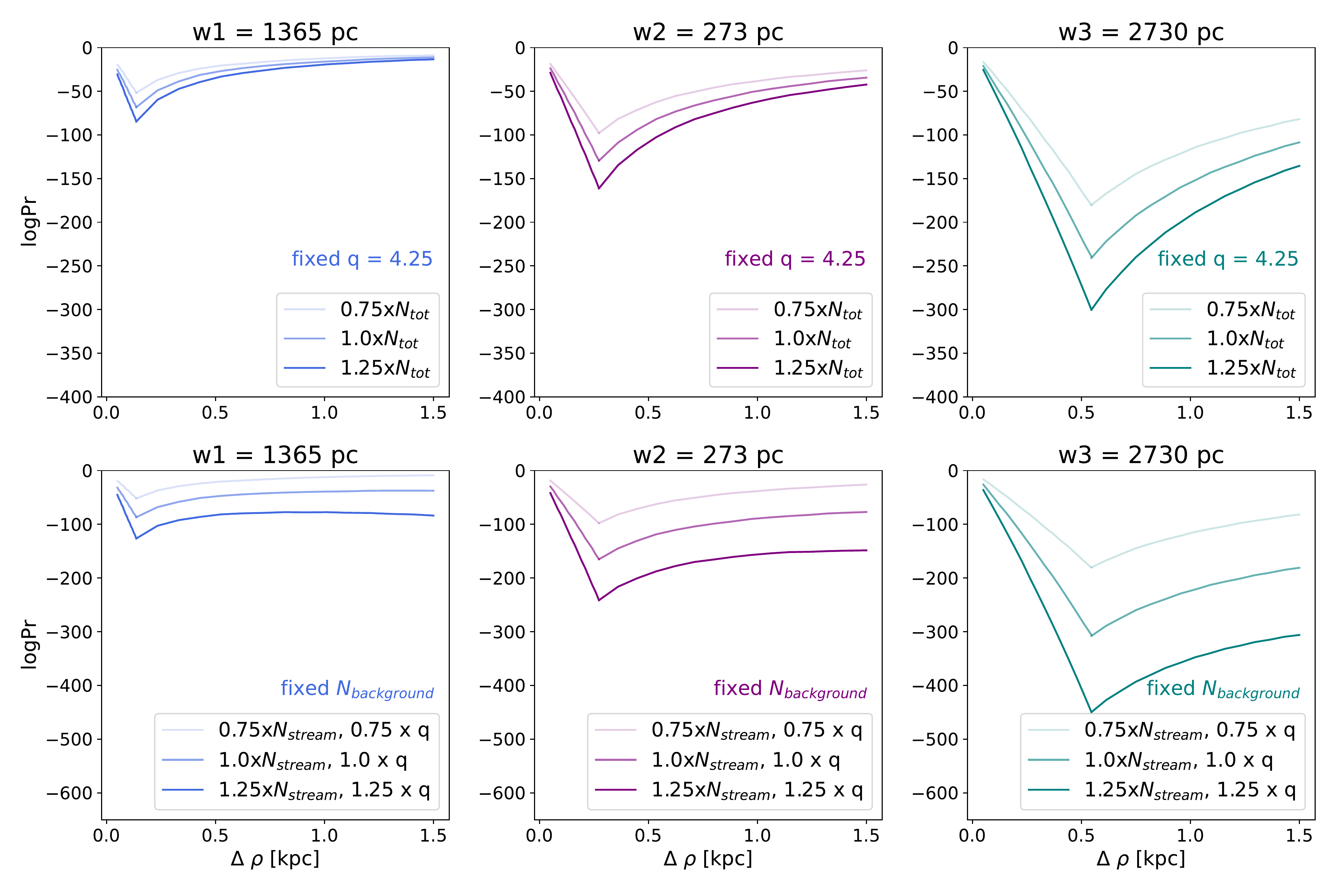}}
\caption{Analytic log$_{10}$Pr values (see Equation \ref{eq:binomial}) versus $\Delta \rho$ for three different streams with three different widths w1 = 136 pc (blue, left), w2 = 273 pc (purple, middle), and w3 = 546 pc (teal, right). ``$q$" is the number density (stars/kpc$^2$) in the stream versus the background.
Top row: the middle magenta line (in the middle panel) is the $10{M}_{\rm Pal 5}$ synthetic stream case where there are $130$ stars in the stream and $879$ stars in the background ($q = 4.25$, see also Figure \ref{fig:drho}). We use this as a starting point and scale our other examples from here. The other magenta lines represent the scenario where we have scaled the number of stars down in both the stream and the background by a factor of 0.75 (lighter) or up by 1.5 (darker). 
The other two panels demonstrate examples with smaller (left) and larger (right) stream widths by a factors of one half and two respectively. To keep $q$ constant, we scaled the number of stars in the streams based on their areas (i.e. there are fewer stars in the stream to the left and more stars in the stream to the right). We note that having more stars in total yields a more significant detection (darker lines), despite constant $q$.
Bottom row: 
Instead of keeping $q$ fixed, here we instead fix the number of stars in the background to be 879 in each panel and  scale the number of stream stars up and down by a factor of 0.75 and 1.25, respectively. A larger number of stars in each stream makes the detections  more significant (smaller log$_{10}$Pr-values). 
}
\label{fig:drho_analytic}
\end{figure*}

We carry out the steps above for three different stream widths (w1 = 136 pc, w2 = 273 pc, and  w3 = 546 pc). We first keep the number density (stars/kpc$^2$) between each stream and the background fixed ($q$) but scale up or down the total number of stars, $N_{\rm tot} = N_{\rm stream} + N_{\rm background}$. To keep the number density ratio, $q$, fixed, we scale up the number of stream stars with for  larger width streams. Thus, there are more total stars in the wider streams.

In the top panel of Figure \ref{fig:drho_analytic}, we show the results for fixed $q$ for the three different widths. As a starting point, we use the stream from the example in Figure \ref{fig:drho}, a $10{M}_{\rm Pal 5}$ stream with 130 stars injected to the PAndAS background at $R_{gc}=55$ kpc with a radius of 5 kpc including 879 stars. We do not change the area of the regions in the following examples. The middle magenta line in Figure \ref{fig:drho_analytic} (top, middle) shows this example (same line as shown in Figure \ref{fig:drho}). Here, $q = 4.25$, and we use this $q$ for all lines in the top panel. The two other panels (top left and right) in Figure \ref{fig:drho_analytic} represent streams with half of the width of the $10{M}_{\rm Pal 5}$ synthetic stream (left: blue) and two times the width of the $10{M}_{\rm Pal 5}$ synthetic stream (right: teal). For constant $q$ but with a higher total number of stars in the streams and backgrounds, the stream is more significantly detected (lower log$_{10}$Pr value). The widest streams with more stars are most significantly detected. 
In each case, the stripe width ($\Delta \rho$) for optimal detection, per construction, is the width of the stream. If the stream had not been assumed to fall perfectly in the center of the input data region, we could have a scenario where the stream is smeared over several $\Delta \rho$-bins (``stripes'') and the signal will be slightly weaker, as is the case in Figure \ref{fig:drho}, for the numerical (dashed line) versus analytic (solid line) example.

In the bottom row of Figure \ref{fig:drho_analytic}, we keep the number of stars in the background fixed ($N_{\rm background} = 879$), but scale up and down the number of stars in the stream. Here, the contrast between the stream and background is getting higher with darker line colors. As expected, the log$_{10}$Pr values become lower (more significant) with higher number densities of stream stars (darker lines). Thus, it is easier to detect streams in low-density backgrounds and streams with more stars (see example in Section \ref{sec:completeness}). 

Based on the analyses here, where we have fixed the region sizes to have $r = 5$ kpc, we conclude the following: 
\begin{itemize}
    \item[1.] A higher number of total stars will lead to a higher significance detection, even with a fixed number density contrast between the stream stars and backgrounds stars (Figure \ref{fig:drho_analytic}, top row).  
    \item[2.] A larger contrast between the stream and background yields a large difference in detection significance (Figure \ref{fig:drho_analytic} bottom row). Thus, it easier to detect streams in the outer halos of galaxies.
    \item[3.] Wider streams yield a more significant detection (Figure \ref{fig:drho_analytic}, left to right). 
\end{itemize}

\section{Completeness test}\label{sec:appendix_completeness}
\begin{figure*}
\centerline{\includegraphics[width=0.7\textwidth]{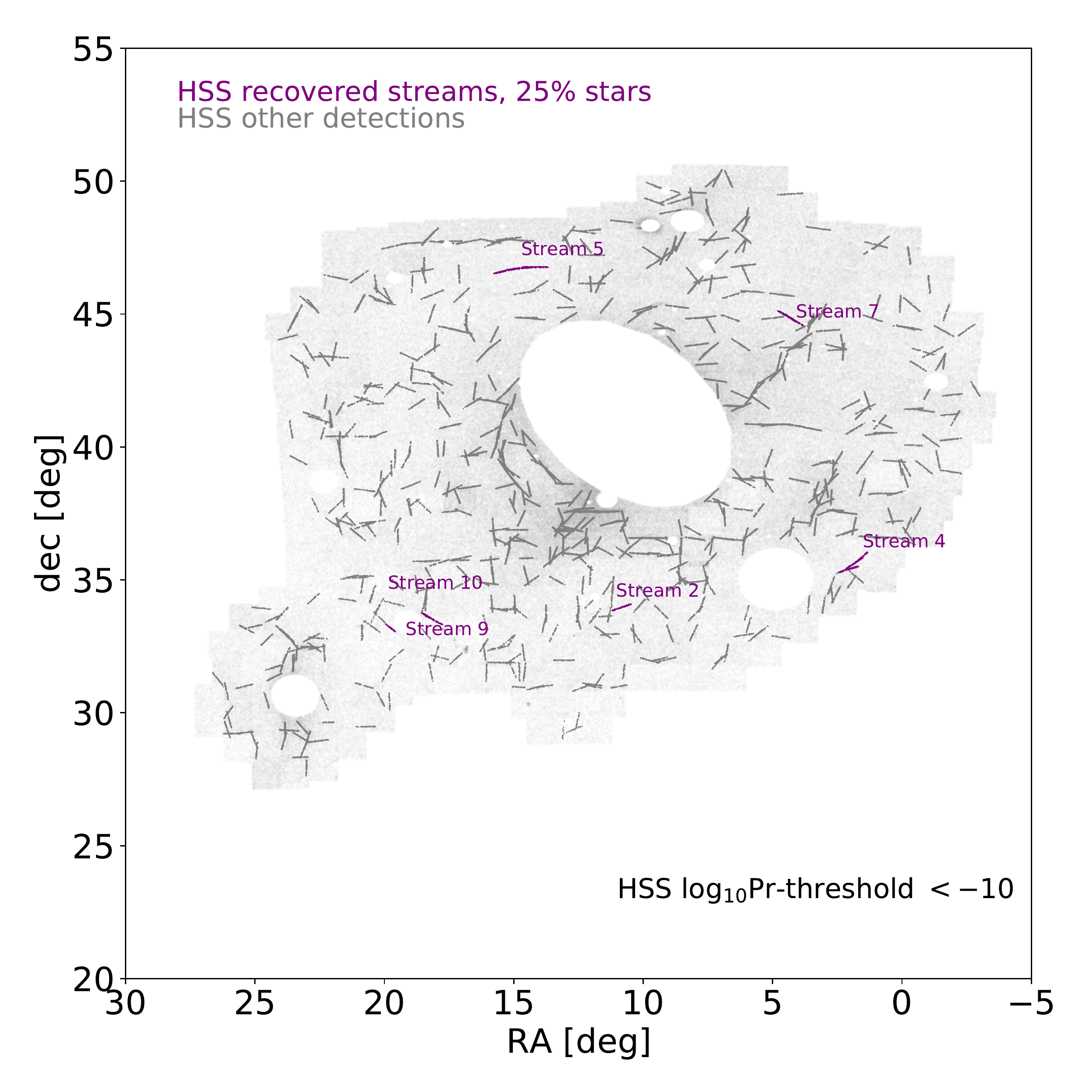}}
\caption{Same as Figure \ref{fig:completness} (right), but including only 25\% of the 311 $10{M}_{\rm Pal 5}$ synthetic stream stars. The purple highlighted stars are stars that were flagged in an \texttt{HSS} run with a Pr-thresh $<-10$ instead of $-15$ as in the case for the 100\%, 75\%, and 50\% runs. Note that the code flags many new linear features (dark gray), and that the code partially recovers five of the ten injected streams (purple). At this Pr-thresh, several of the recovered features trace out the artifacts at 0 and 90 $\deg$ from the CFHT $1\deg \times1\deg$ pointings and many trace out part of the known, wider dwarf debris features.}
\label{fig:completness_logPr10_25p}
\end{figure*}

We highlight the findings of the \texttt{HSS} run with ten injected $10{M}_{\rm Pal 5}$ synthetic streams including only 25\% of the initial stars (see also Section \ref{sec:completeness}). Note that these streams are not visible by eye in the data (Figure \ref{fig:completness}, left panel and lower right panel). As the \texttt{HSS} did not recover any of the synthetic streams if we ran the code with Pr-thresh $<-15$, in this run we used Pr-thresh $<-10$. In Figure \ref{fig:completness_logPr10_25p}, purple streaks highlight the stars that were recovered from the injected streams. Five of the streams are partially recovered, but streams 3 and 10 were only flagged in one region. The dark gray streaks highlight the other linear features that the \texttt{HSS} flagged in this run. Several of these flagged detections appear to trace out the linear artifacts (at 0 $\deg$ and 90 $\deg$) that are present in the data due to the CFHT $1\deg \times1\deg$ FOV, and many are in the close vicinity of the known dwarf debris. Additionally, several unknown GC candidates were flagged. It is possible that some of these are true stream candidates, but based on our tests in Section \ref{sec:optimize}, they are likely noise.

\section{Searching for Pal 5--like streams}\label{sec:appendix_pal5}
\begin{figure*}
\centerline{\includegraphics[width=0.7\textwidth]{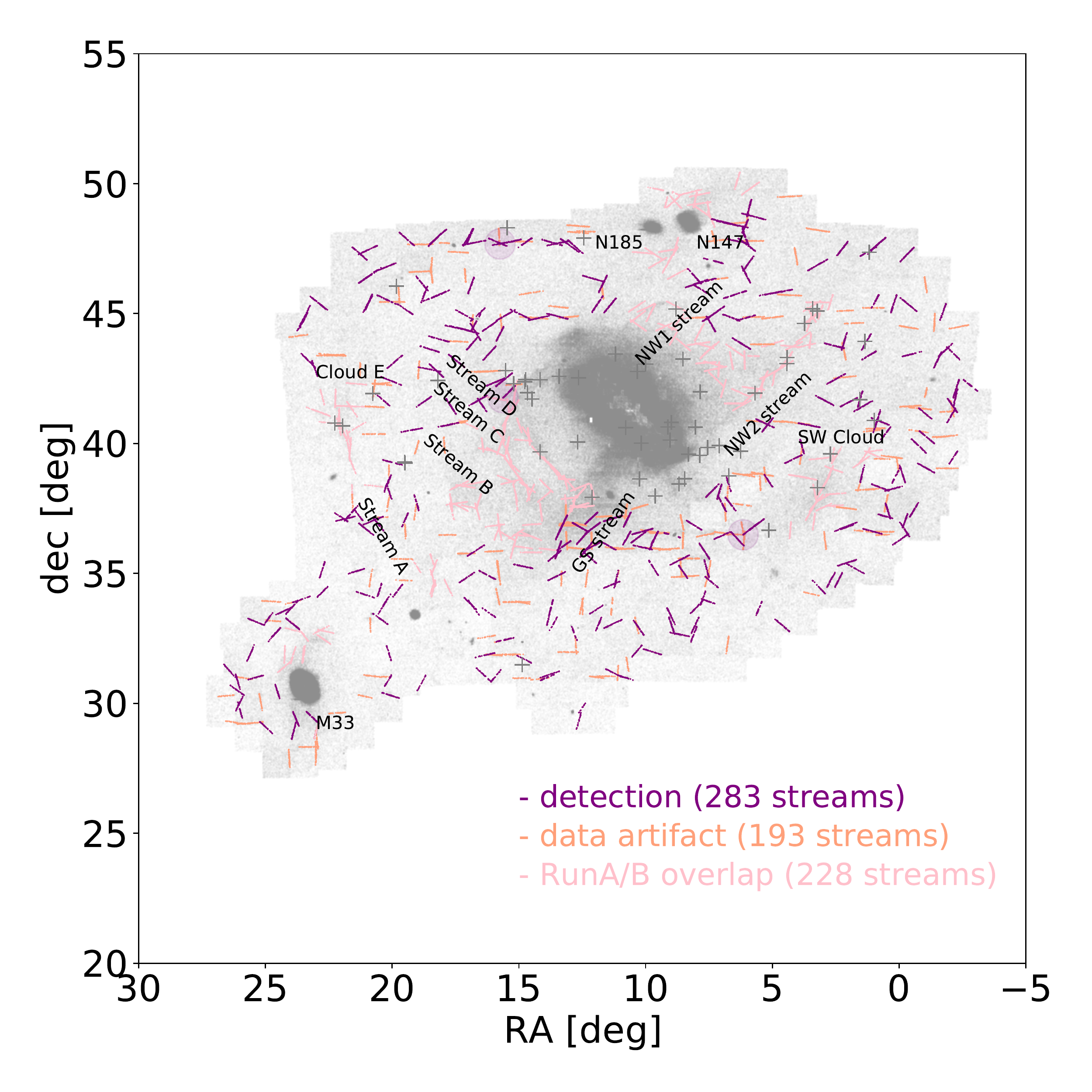}}
\caption{Flagged \texttt{HSS} candidates from a blind run on 2766 overlapping PAndAS data regions (with [Fe/H] $<-1$) and a radius of $r_{\rm angular} = 0.365~ \deg$ with search parameters set to $\Delta \rho = 0.3$ kpc, Pr-thresh $<-10$ to search for Pal 5--like streams. The known dwarfs, M31, and known GC were masked out in this run. The \texttt{HSS} flags 705 candidates streams, where 283 streams (purple) are potential new GC candidate streams, and where 193 of these are likely data artifacts (salmon) as $\theta = 0$ or $90$ $\deg$ and trace out the CFHT $1\deg \times1$ $\deg$ fields. The remaining 228 of the flagged streams fall within 0.5 $\deg$ of known dwarf streams found in \texttt{RunA}/\texttt{RunB} (pink). The gray ``+" markers indicate the location of outer halo GCs in the PAndAS data \citep[][]{huxor14}.
}
\label{fig:2pal5}

\end{figure*}

The \texttt{HSS} is capable of detecting lower-mass, shorter $2 {M}_{\rm Pal 5}$--like streams (length = 12 kpc, width = 127 pc, nstars = 68) in the PAndAS data with [Fe/H] $<-1$. However, at the limiting magnitudes of PAndAS, the \texttt{HSS} will also pick up a lot of noise. In Figure \ref{fig:2pal5}, we show the result of the \texttt{HSS} run with $\Delta\rho = 0.3$ kpc and Pr-thresh $= -10$, where the \texttt{HSS} flags 705 detections total. While some of the purple streams might indeed be real Pal 5 like streams, we cannot confirm this with the data in hand. We would instead need either deeper follow-ups or future observations with Roman (see Section \ref{tab:roman}). 

\end{document}